%% file: 00-Main.tex
\documentclass[conference]{IEEEtran}
\IEEEoverridecommandlockouts

\input{preamble}
\input{macros}

\def\BibTeX{{\rm B\kern-.05em{\sc i\kern-.025em b}\kern-.08em
    T\kern-.1667em\lower.7ex\hbox{E}\kern-.125emX}}
\begin{document}

\pdfpagewidth=8.5in
\pdfpageheight=11in

\newcommand{\iscasubmissionnumber}{3547}


\title{AXLE: Coordinated Offloading with Asynchronous Back-Streaming in Computational Memory Systems}


\author{\IEEEauthorblockN{Suyeon Lee}
\IEEEauthorblockA{\textit{School of Computer Science} \\
\textit{Georgia Institute of Technology}\\
Atlanta, USA \\
sylee0506@gatech.edu}
\and
\IEEEauthorblockN{Kangkyu Park}
\IEEEauthorblockA{\textit{Memory System Research} \\
\textit{SK hynix Inc.}\\
Icheon, South Korea \\
kangkyu.park@sk.com}
\and
\IEEEauthorblockN{Kwangsik Shin}
\IEEEauthorblockA{\textit{Memory System Research} \\
\textit{SK hynix Inc.}\\
Icheon, South Korea \\
kwangsik.shin@sk.com}
\and
\IEEEauthorblockN{Ada Gavrilovska}
\IEEEauthorblockA{\textit{School of Computer Science} \\
\textit{Georgia Institute of Technology}\\
Atlanta, USA \\
ada@cc.gatech.edu}
}

\maketitle


\begin{abstract}

CXL-based Computational Memory (CCM) enables near-memory processing within expanded remote memory, 
\sheph{offering} opportunities to address data movement costs 
\sheph{in} disaggregated memory systems and to accelerate overall performance. However, existing 
offloading mechanisms do not fully leverage the trade-offs of different offload models based on different CXL protocols.
This work first examines these tradeoffs and 
their impact on end-to-end performance and system efficiency 
\sheph{for workloads with diverse data and computation characteristics.}
We propose Asynchronous Back-Streaming, \sheph{a new offloading} protocol 
\sheph{that coordinates CXL.io and CXL.mem to enable result back-streaming and asynchronous pipelining across CCM and host tasks.}
We \sheph{further} design \system, a system that realizes 
\sheph{this protocol with lightweight host-CCM interaction.}
Overall, \system reduces end-to-end runtime by up to \iscaRev{50.14\%}, \iscaRev{reduces} CCM and host idle times by \iscaRev{an average of} \iscaRev{14.53$\times$} and \iscaRev{3.93$\times$}, 
\sheph{respectively, and achieves up to 6$\times$ reduction in host core stall time.}

\end{abstract}


\input{01-Introduction}
\input{02-Background}
\input{03-Motivation}
\input{04-System}
\input{05-Evaluation}
\input{06-Related}
\input{07-Discussion}
\input{08-Conclusion}

\section*{Acknowledgements}
We thank the anonymous reviewers and shepherd for their constructive feedback. This research was partially supported by the Intel Center for Transformative Server Architecture, and the Center for Processing with Intelligent Storage and Memory (PRISM), a JUMP 2.0 joint program by the Semiconductor Research Corporation and DARPA.


\newpage
\bibliographystyle{IEEEtranS}
\bibliography{10-Bibliography}


\end{document}

%% file: preamble.tex
\usepackage{gensymb}
\usepackage{booktabs}  
\usepackage{pbox}
\usepackage{subfigure}
\usepackage{epsfig,endnotes}
\usepackage{epstopdf}
\usepackage{color}
\usepackage{graphicx}
\usepackage{xspace}
\usepackage{flushend}  
\usepackage{enumitem}
\usepackage{multirow}
\usepackage{tikz}
\usepackage[ruled,vlined,linesnumbered,noend]{algorithm2e}

\usepackage{url}

\usepackage{breakurl}
\usepackage{hyperref}
\usepackage{mathtools}

\usepackage{xspace}
\usepackage{cleveref}
\usepackage{comment}

\usepackage{amsmath}
\usepackage[framemethod=TikZ]{mdframed}
\usepackage[normalem]{ulem}

\usepackage{hyperref}
\usepackage{soul}
\usepackage{tabularx}

\usepackage{algorithmic}

\usepackage[table]{xcolor}
\usepackage{pifont}
\newcommand{\cmark}{\ding{51}}%
\newcommand{\xmark}{\ding{55}}%

\usepackage{lipsum}

\usepackage[noframe]{showframe}
\usepackage{framed}
\renewenvironment{shaded}{%
  \MakeFramed{\advance\hsize-\width \FrameRestore\FrameRestore}}%
 {\endMakeFramed}
\definecolor{shadecolor}{gray}{0.9}

\usepackage{booktabs}
\usepackage{multirow}
\usepackage{array}

\usepackage{textcomp}

%% file: macros.tex
\newif\ifcomment
\commenttrue

\newif\ifwatermark
\watermarkfalse

\ifwatermark
      \usepackage{draftwatermark}
      \SetWatermarkScale{.7}
      \SetWatermarkText{\shortstack{In Submission:\\Do Not Distribute}}
      \SetWatermarkColor[rgb]{1,0.88,0.88}
\fi

\definecolor{green}{rgb}{0.35, 0.71, 0.3}

\ifcomment
    \newcounter{DNumberOfComments}
    \stepcounter{DNumberOfComments}
    \newcommand{\dnote}[1]{\textcolor{blue}{\small \bf [DZ\#\arabic{DNumberOfComments}\stepcounter{DNumberOfComments}: #1]}}

    \newcounter{KDNumberOfComments}
    \stepcounter{KDNumberOfComments}
    \newcommand{\suyeon}[1]{\textcolor{magenta}{\small \bf [SY\#\arabic{KDNumberOfComments}\stepcounter{KDNumberOfComments}: #1]}}
    \newcommand{\ada}[1]{\textcolor{blue}{\small \bf [AG\#\arabic{KDNumberOfComments}\stepcounter{KDNumberOfComments}: #1]}}
\else
    \newcommand\dnote[1]{}    
    \newcommand\kdnote[1]{}
    \newcommand\synote[1]{}
    \newcommand\suyeon[1]{}
    \newcommand\ada[1]{}
\fi

\smallskip
\smallskip

\newcommand{\ra}[1]{\renewcommand{\arraystretch}{#1}}

\newcommand{\system}{{AXLE}\xspace}
\newcommand{\baseline}{{M$^2$NDP}\xspace}

\usepackage{calc}
\newlength\myheight
\newlength\mydepth
\settototalheight\myheight{Xygp}
\newcommand{\heading}[1]{\noindent\textbf{#1.}}

\newcommand{\headingit}[1]{\noindent\textbf{\textit{#1.}}}

\mdfsetup{skipabove=\topskip,skipbelow=\topskip}
\newcounter{observ}[section]

\DeclareRobustCommand\circledchar[2]{\tikz[baseline=(char.base)]{
		\node[shape=circle,draw=none,color=white,fill=black,inner sep=#2] (char) {#1};}}

\crefformat{section}{\S#2#1#3}
\crefrangeformat{figure}{Figures~#3#1#4--#5#2#6}

\newcommand{\cred}{}
\newcommand{\iscaRev}{}
\newcommand{\sheph}{}

\DeclareRobustCommand\circledchar[2]{\tikz[baseline=(char.base)]{
		\node[shape=circle,draw=none,color=white,fill=black,inner sep=#2] (char) {#1};}}

%% file: 01-Introduction.tex
\section{Introduction}\label{sec:intro}




High demands for reducing data movement bottlenecks and for solving memory capacity problems have paved the way for memory disaggregation in recent datacenters~\cite{infiniSwap, vmwareRemoteMem, vmwareMemDis}.
Compute eXpress Link (CXL)~\cite{cxlSpec, intelCXL, demystifyingCXL} has emerged as a promising interconnection technology for efficient, high-performance disaggregated memory systems~\cite{directCXL, pond, tpp}.
However, as the performance gap between processing units and memory grows, it becomes challenging to hide data movement from the critical path, leaving memory and fabric major bottlenecks.
This makes the case for \iscaRev{adopting} emerging CXL-based computational memory, \textbf{CCM}, which incorporates a \iscaRev{processing-near-memory (PNM)} unit in a remote memory.


One of the main approaches to integrate the emerging CCM technology in existing systems is to partially offload memory-intensive operations within the applications (\Cref{tab:back_workloadOffloading}).
Prior works in this domain focus on \textit{which} operation to  offload~\cite{m2ndp, udon, CMS, CXL-ANNS, neuPIMs, CLAY, PIFS-REC, BEACON, StRoM, NearPM, monde}\iscaRev{,~\cite{grudon}}. This leads to application-specific \iscaRev{PNM} solutions which accelerate partial tasks.
Such scenarios have been well validated across a wide range of applications, given the diversity of application tasks and the different processing capabilities between the host and the CCM module.

\begin{figure}[t]
    \centering
    \includegraphics[width=0.88\linewidth]{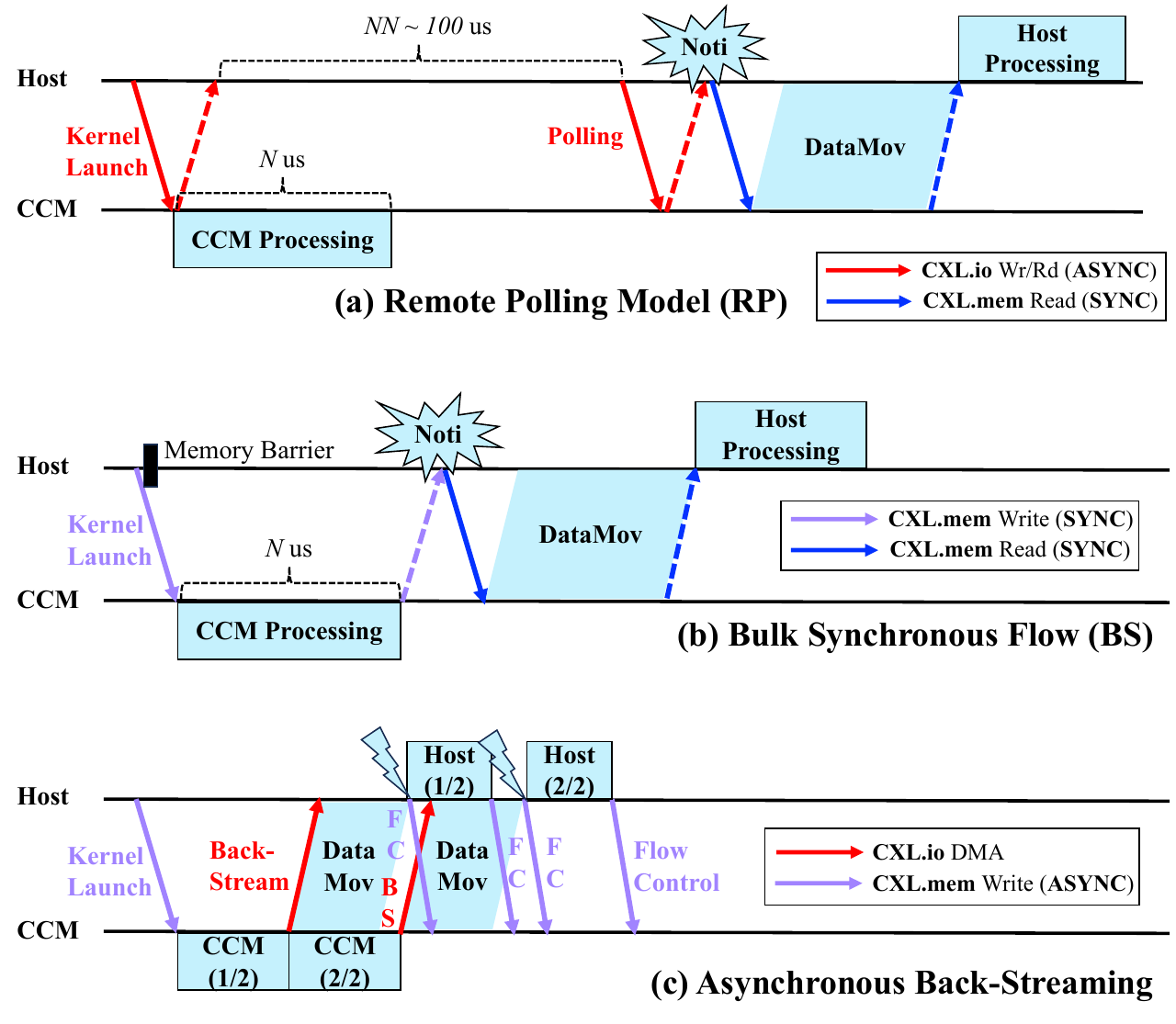}
    \caption{\cred{Simplified view of existing CCM partial offloading mechanisms (a, b) and the mechanism proposed in this work (c). Dotted lines represent ACKs/responses for the corresponding memory requests, omitted in (c) as they are unnecessary under our fully asynchronous interaction.}}
    \label{fig:intro_models_overview}
    \vspace{-0.15in}
\end{figure}

Unlike most existing studies, this work focuses on \textit{how} to perform the partial offload.
This is not trivial, since CCM can be used both as a device and as memory.
\cred{\Cref{fig:intro_models_overview} illustrates existing partial offloading mechanisms and how our new protocol improves end-to-end \iscaRev{runtime}.}
With a traditional device-centric view, most of the previous systems rely on CXL.io messages for task offloading and a remote polling mechanism \cred{(\Cref{fig:intro_models_overview}(a))}.
This enables asynchronous remote task execution, however, it is not suitable for fine-grained offloading due to the high CXL.io-based remote polling overheads.
Recently, \baseline~\cite{m2ndp} proposed a  CCM architecture that views CCM from a memory-centric perspective.
It supports low-overhead 
task offloading by utilizing CXL.mem-based host-CCM communication \cred{(\Cref{fig:intro_models_overview}(b))}.
This reduces the offloading overhead and enables fine-grained task offloading.
However, the underlying CXL.mem memory semantics introduce bulk-synchronous data loads and cause the host CPU to be idled during CCM task processing. 
Our evaluation using a graph analytics benchmark shows that up to 98\% of the host and approximately 50\% of the CCM remain idle during the total runtime (\Cref{subsec:motiv_idleTimes}). 
Therefore, existing operation offloading mechanisms are 
limited by the CXL.io vs.~CXL.mem host-CCM communication models they use. 
%
In addition, it is not sufficient to consider only the speed 
of invoking and executing offloaded operations.
Rather, the focus should be on the end-to-end execution of the application pipeline, which integrates both host and CCM computations while coordinating the exchange of data and commands between them.



To address these challenges, we propose a novel \textbf{asynchronous back-streaming} protocol \cred{(\Cref{fig:intro_models_overview}(c))} for host–CCM coordination, along with a system, \textbf{\system}, that implements it.
The asynchronous back-streaming protocol enables continuous overlap of different components, thereby minimizing end-to-end runtime and resource idle times in the host–CCM interaction pipeline.
Its core concept is to allow the CXL device to trigger reverse data streaming from the remote to the local memory, coupled with asynchronous pipelining of
\sheph{upstream CCM,} data movement and
\sheph{downstream host tasks.}
The new protocol realizes the offloading mechanism by leveraging the strengths of both the CXL.io and CXL.mem protocols: 
CXL.io DMA 
\textit{asynchronously sends partial result} from the CCM to the host, in contrast to prior models that rely on \textit{full synchronous result loads} triggered by host processing units.
To launch the offloading kernel and manage the DMA region on the local host, \system uses CXL.mem memory requests for control messages, 
thereby retaining low protocol overheads in the critical path.

\system is a 
system that \sheph{realizes} 
the asynchronous back-streaming model. 
To enable rapid and efficient notification of partial result availability, \system relocates the polling point to the local host region, partitions the DMA region into two ring buffers for metadata and payload for lightweight polling, and supports fully asynchronous CCM–host communication.
DMA-based result streaming delivers partial result data in advance to the local region, enabling the host processing units to access the data locally during task execution.
In addition, \system supports out-of-order \iscaRev{(OoO)} streaming, providing an interface \iscaRev{to} flexibly integrate with existing CCM and host parallel task schedulers~\cite{CXL-ANNS, neuPIMs, PIFS-REC, BEACON}\iscaRev{,~\cite{grudon}} while keeping them isolated, without requiring synchronization of task execution orders.

We compare asynchronous back-streaming and \system against the two existing partial offloading mechanisms: based on remote polling (RP) vs.~bulk synchronous (BS) flow.
Both baselines are implemented on top of the state-of-the-art CCM architecture, \baseline.
\cred{We also implement an AXLE variant that adopts a different design choice as an additional baseline.}
We evaluate several workloads with different data movement, CCM and host \iscaRev{runtimes}.
Our results show that \system improves end-to-end performance by up to \iscaRev{50.14\%} compared to RP, 
and by up to \iscaRev{48.88\%} compared to BS. 
Additionally, \system reduces \sheph{application-level} CCM idle time \iscaRev{by an} average \iscaRev{of} \iscaRev{13.99$\times$} and \iscaRev{14.53$\times$} relative to 
RP and BS, 
respectively, and reduces host idle time \iscaRev{by an} average \iscaRev{of} \iscaRev{3.93$\times$} and \iscaRev{3.85$\times$}. 
\sheph{Furthermore, AXLE reduces host core stall time by up to 6$\times$, improving host core utilization.}

This paper makes the following contributions:
\begin{itemize}[noitemsep, topsep=0pt]
    \item We present the duality of CCM from device-centric and memory-centric perspectives, highlighting unexploited trade-offs arising from the underlying mechanisms and CXL protocols (\Cref{sec:motiv}).
    We emphasize an end-to-end pipeline perspective of CCM systems, showing how diverse workload characteristics can lead to suboptimal performance and idle times at both the host and the CCM. 
    \item We propose a new protocol \iscaRev{for CCM offloading,} asynchronous back-stream\-ing, 
    \sheph{which uniquely coordinates CXL.io DMA and CXL.mem to enable} 
    continuous overlap of components, thereby reducing end-to-end runtime and minimizing idle times in the host–CCM interaction pipeline (\Cref{sec:system}).
    \item We design \system, which embodies asynchronous back-streaming as its offloading mechanism. \system supports lightweight host pipelining, proactive back-stream\-ing of data, and an \iscaRev{OoO} streaming interface that 
    increases data movement parallelism and performance, 
    \cred{while ensuring ordering correctness}
    (\Cref{sec:system}).
    \item We evaluate \system through detailed simulations and compare its performance against \baseline under \cred{various} partial offloading mechanisms. 
    \sheph{Across diverse workloads, \system provides significant improvements in end-to-end runtime, and host and CCM efficiency.}
    (\Cref{sec:eval}).
\end{itemize}

%% file: 02-Background.tex


\section{CCM: CXL-based Computational Memory}\label{sec:background}



\heading{Model} Compute eXpress Link (CXL) is a PCIe-based interconnect that provides cache-coherent access to remote devices using memory semantics~\cite{cxlSpec, intelCXL, demystifyingCXL}.
CXL defines three protocols: CXL.io, CXL.cache, and CXL.mem.
It allows composing different types of devices by combining protocols.
\sheph{Type 1 devices (e.g., smart NICs without device memory) combine CXL.io and CXL.cache for cache-coherent access to host memory, and Type 2 devices (e.g., GPUs) further add CXL.mem to expose their own local memory to the host.}
A common use case for CXL are Type 3 devices~\cite{directCXL, demystifyingCXL}, which mix the CXL.io and CXL.mem protocols to expand memory capacity beyond local servers.
CXL.io is a drop-in replacement for the PCIe protocol, whereas CXL.mem enables byte-addressable
access to 
expanded memory regions 
using typical \texttt{load} and \texttt{store} instructions.

\begin{table}[t]
    \centering
    \caption{Target application benchmarks and the memory-intensive operations they offload to CCM.}
    \bgroup
    \def\arraystretch{1.05}
    \begin{tabular}{|p{0.13\textwidth}|p{0.3\textwidth}|}
        \hline
        \textbf{Workload} & \textbf{Offloaded Function} \\ 
        \hline
        OLAP/OLTP & Filtering (e.g., within \texttt{SELECT})~\cite{m2ndp} \\
        \hline
        Graph Analytics & Edge traversal $\rightarrow$ Vertex update~\iscaRev{\cite{grudon}} \\
        \hline
        KNN/ANN & Vector distance calculation~\cite{udon, CMS, CXL-ANNS} \\
        \hline
        LLM Inference & Attention block~\cite{neuPIMs} \\
        \hline
        \iscaRev{DLRM} & \iscaRev{Embedding table lookup $\rightarrow$ Sparse Length Sum (SLS)~\cite{m2ndp, CLAY, PIFS-REC}} \\
        \hline
    \end{tabular}
    \egroup
    \label{tab:back_workloadOffloading}
\vspace{-0.15in}
\end{table}

CCM is an emerging technology that incorporates computing resources on top of a CXL Type 3 device.
\sheph{We further discuss the implications of utilizing Type 3 devices for CCM in \Cref{sec:discussion}.}
Its computing capability is limited in terms of processing speed and power, or auxiliary resources such as cache. 
However, the embedded CXL Type 3 devices offer high memory performance with respect to the CCM-local compute resources.
Therefore, the primary purpose of CCM is to enable \iscaRev{PNM} for memory-intensive tasks~\cite{m2ndp}\iscaRev{,~\cite{grudon}}.
One of the common use cases is to \textit{partially offload} memory-intensive operations within the applications; we illustrate representative examples in \Cref{tab:back_workloadOffloading}.


\heading{Real Prototypes} Real hardware CCM prototypes have been proposed by industry~\cite{CMS, CXL-ANNS, CXL-PNM}\iscaRev{,~\cite{COSMOS}} and utilized in prior research.
Commonly, these devices rely on application-specific integrated circuits (ASICs) and hardwired primitive function logics (PFLs).
For example, the specific device considered in this work
is an add-in card custom-developed board with a CXL memory controller and \iscaRev{PNM} engine integrated into an FPGA.
In the initial prototype of the real hardware, the \iscaRev{PNM} engine was implemented \iscaRev{with PFLs} designed to support a specific single application such as KNN. 
This approach aimed to achieve optimized acceleration for targeted applications, resulting in impressive performance improvements.

\begin{figure}[t]
    \centering
    \includegraphics[width=0.7\linewidth]{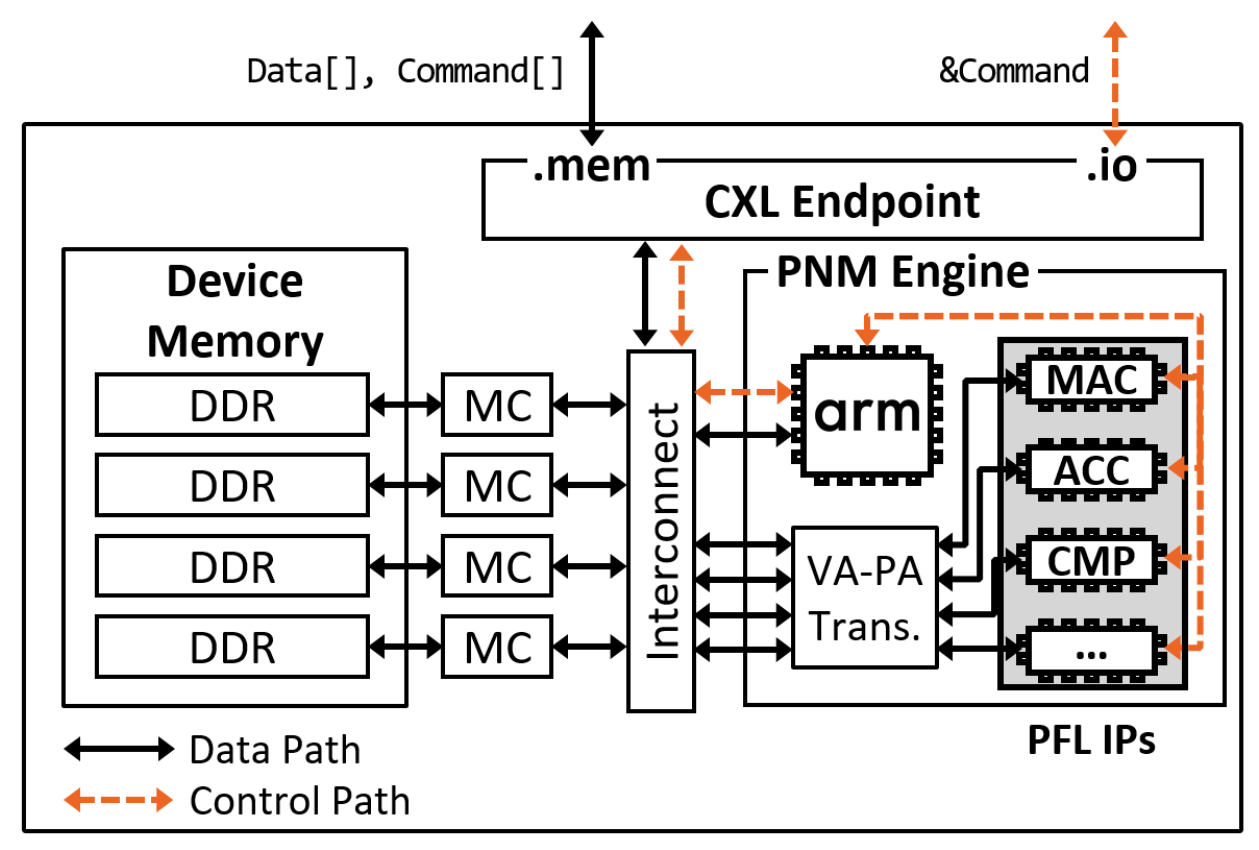}
    \caption{Block diagram of a real prototype of CCM device. The device appears as an endpoint that supports the CXL protocols and memory expansion. It integrates both FPGA-based hardwired PFLs and single general-purpose core.}
    \label{fig:back_cmmAx_arch}
    \vspace{-0.1in}
\end{figure}

As shown in \Cref{fig:back_cmmAx_arch}, the hardware prototype is built around a Xilinx Versal (VP1502) FPGA chip with DRAM mounted across four DIMM slots.
The \iscaRev{PNM} engine provides PFL hardware IP, such as MAC (Multiply Accumulate), ACC (Accumulate), and CMP (Compare), as essential processing blocks for functionalities including numeric/string filtering, vector distance calculation, etc.
Additionally, the use of a Cortex-A72 ARM processor as a general-purpose computational unit offers flexibility for adding new operations.

\heading{Simulation Infrastructure}
The state-of-the-art CCM architecture is \baseline, which provides a design of a low overhead and low cost general-purpose CCM~\cite{m2ndp}.
\baseline achieves remarkable speedups and energy savings across a variety of workloads, compared to baseline CPU/GPU hosts with CXL memory expansion without \iscaRev{PNM}.
The \baseline testbed is based on its own open-source simulator~\cite{m2ndpSim}, a combination of Ramulator~\cite{ramulator} as CXL memory devices and BookSim2~\cite{booksim} as CXL interconnect protocols.

As shown in \Cref{fig:back_cmmAx_arch}, these prototypes \cred{largely rely on} specific hardwired logic, \cred{making them unsuitable as general-purpose devices for diverse workloads.}
\cred{In addition, current hardware prototypes often experience high latency due to immature CXL IP implementations.
As a result, both the architectural components and achievable performance of existing hardware still fall short of what the \baseline architecture envisions (\Cref{subsec:system_overview}), making proper evaluation of the new data and control planes infeasible.}
Instead, the \baseline simulator offers ease of access, flexibility to support diverse workloads, and a high-performance CCM \cred{model}.
\cred{For these reasons,} we use the \cred{validated} \baseline simulator as our primary testbed.
\cred{This simulation-based research serves as a preparatory step toward realizing and validating the new data and control planes on an upcoming ASIC-based CCM device.} 


%% file: 03-Motivation.tex
\definecolor{Gray}{gray}{0.9}
\begin{table}[t]
\centering
\caption{Summary of trade-offs arising from the duality of CCM system architectures, and benefits of asynchronous back-streaming in leveraging the strengths of both modes.}
\scalebox{0.9}{
    \begin{tabular}{|l||c|c|c|c|} 
    \hline
    Partial Offloading Mechanism & \begin{tabular}{@{}c@{}}Fine- \\ grained \\ Offloading\end{tabular} & \begin{tabular}{@{}c@{}}CXL \\ Protocol \\ Overhead\end{tabular} & \begin{tabular}{@{}c@{}}Async \\ Execution \end{tabular} \\ \hline\hline
    Remote Polling~\cite{CMS, CXL-ANNS} & \xmark & High & \cmark \\ \hline
    Bulk Synchronous Flow~\cite{m2ndp} & \cmark & Low & \xmark \\ \hline
    \rowcolor{Gray}
    Asynchronous Back-Streaming & \cmark & Low (Hidden) & \cmark \\ \hline
    \end{tabular}
}
\vspace{-0.15in}
\label{tab:motiv_mechnisms}
\end{table}

\section{Motivation}\label{sec:motiv}



\subsection{Duality of Computational Memory}\label{subsec:motiv_duality}


Given that CCM integrates both compute \textit{and} memory, it can be perceived from two perspectives: \textit{device-centric} view and \textit{memory-centric} view.

Device-centric view~\cite{CMS, CXL-ANNS} assumes CCM is viewed as an accelerator, and operation offloading is performed \iscaRev{primarily} via CXL.io.
It uses CXL.io for various steps in host-CCM communications required to offload the function through a remote mailbox access (MMIO register on the CXL device).
A key mechanism 
in this setting is  \textit{remote polling} \cred{(RP; \Cref{fig:intro_models_overview}(a))}. 
The local host needs to \iscaRev{initially} write the application kernel descriptor to the CXL memory \iscaRev{via CXL.mem, then use CXL.io to} \iscaRev{\textit{(\underline{1})}} enqueue the offloading command, and \iscaRev{\textit{(\underline{2$\sim$n})}} start polling the mailbox to check if the remote kernel is completed.
When the CXL firmware writes the completion descriptor in the mailbox, the host can acknowledge it via polling response.
Then, \textit{(\underline{n+1})} the host sends the final CXL.io message to  dequeue the offloading command.
Lastly, the host sends a CXL.mem message to load the offloading results before processing any dependent host kernel.

The CXL.io-based \cred{interactions are} 
asynchronous, and provide an opportunity to avoid blocking the host processing due to remote kernel execution.
The main drawback of the \cred{RP} model is that it cannot support offloading of fine-grained tasks which take on the order of microseconds processing time~\cite{m2ndp}.
Its mechanism requires \cred{remote} polling between \cred{the host and the device,} 
\cred{where its} polling interval is up to 100 microseconds in a real-hardware setup.
Moreover, it adds up CXL.io round-trip time~\cite{cxlSpec, intelCXL} to poll the remote region.
These CXL.io-based message exchanges cannot be hidden within the pipeline.
\cred{As a result, remote polling inherently limits the efficiency of host–CCM interaction and becomes a bottleneck when offloading fine-grained kernels.}

Meanwhile, in a memory-centric view, CCM is accessed as a memory device. It supports operation offloading via CXL.mem, where the mechanism implies \textit{bulk synchronous flow} \cred{(BS; \Cref{fig:intro_models_overview}(b))}.
To invoke remote functions via 
\sheph{memory operations,}
\baseline~\cite{m2ndp} proposes several hardware features.
A custom packet filter on the CXL memory controller allows the hardware to differentiate between  basic memory operations and remote kernel launch.
Thus, the host can offload a function simply by issuing a single CXL.mem store operation of the kernel information to the specific remote address range.
In this case, a synchronous CXL.mem store response indicates the remote kernel completion.
To block other memory operations until the response arrives back at the host, the CXL memory controller also relies on \cred{memory barriers.} 

\begin{figure}[t]
    \centering
    \subfigure[Compute-heavy Tasks]{
        \label{fig:motiv_baselines_bulkTasks}
        \includegraphics[width=0.453\linewidth, trim={0 0 0 0cm},clip]{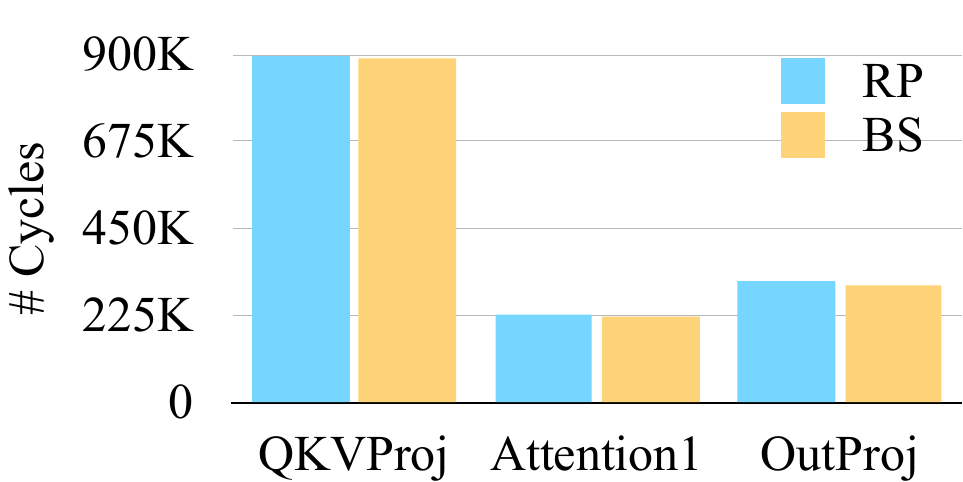}
    }
    \hfill
    \subfigure[Lightweight Tasks]{
        \label{fig:motiv_baselines_fineGrainedTasks}
        \includegraphics[width=0.451\linewidth,trim={0 0 0 0cm},clip]{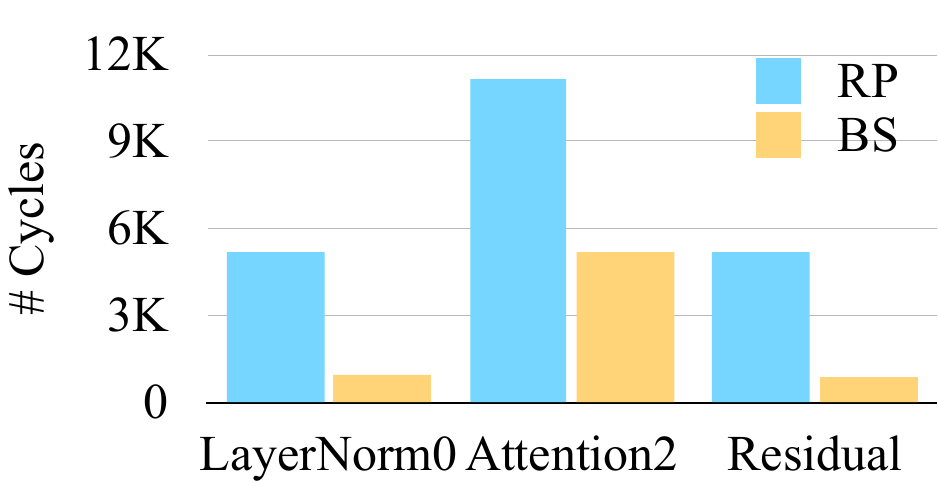}
    }
    \caption{Kernels of the attention block in LLM inference, exhibiting different characteristics under the OPT-2.7B model with a token size of 1K.}
    \label{fig:motiv_baselines}
    \vspace{-0.1in}
\end{figure}

The \cred{BS} model effectively solves the existing problems of the \cred{RP} model.
\Cref{fig:motiv_baselines} demonstrates the case of running multiple kernels of the attention block within LLM inference, using both models. 
The kernels are based on the \baseline benchmark~\cite{m2ndpSim} and the attention block execution order is \texttt{LayerNorm0}, \texttt{QKVProj}, \texttt{Attention1}, \texttt{Attention2}, \texttt{OutProj}, and \texttt{Residual}.
Among them, half are 
computationally heavy
tasks as shown in \Cref{fig:motiv_baselines_bulkTasks}, where the number of cycles spent to run QKVProj is up to 897K when using RP. 
In these cases, the BS model 
results in similar number of cycles, for example, running QKVProj on top of it takes 888K.
In contrast, \Cref{fig:motiv_baselines_fineGrainedTasks} shows the case of running the 
more lightweight tasks whose  number of execution cycles is much less than the heavy tasks.
The BS model 
incurs significantly fewer cycles to execute these tasks:
only 16.7\% of the cycle count when using \cred{the} RP. 
This means that the BS model 
largely reflects the pure \iscaRev{runtime} of the kernel, whereas the RP 
model suffers from long polling intervals and associated overheads, which significantly increase the overall runtime when offloading fine-grained tasks. 

The use of CXL.mem \cred{enables} both fine-grained and coarse-grained offloading without the limitations \cred{imposed by remote} polling over the CXL link \cred{and its associated overheads.}
However, since the mechanism relies on synchronous CXL.mem operations to execute remote kernels, the host processing unit \cred{stalls} until the remote \cred{execution} completes \cred{and the results are loaded}.
%
%
\Cref{tab:motiv_mechnisms} summarizes the trade-offs \cred{stemming from} the duality of CCM system architectures
\cred{and highlights how our proposed} \textit{asynchronous back-streaming} model \cred{(\Cref{fig:intro_models_overview}(c))} leverages the strengths of \cred{both} modes
\cred{to} support efficient, general-purpose CCM systems.


\vspace{-0.1in}
\begin{shaded}
\heading{Observation \#1: Trade-offs in duality of CCM}
The device-centric view relies on remote polling mechanism and allows asynchronous operation offloading. The memory-centric view is based on
bulk synchronous flow and enables fine-grained offloading.
\sheph{By treating CCM as either a device or memory alone, existing mechanisms miss the opportunity to combine the strengths of both CXL.io and CXL.mem.}
\end{shaded}

\subsection{Workload Considerations}\label{subsec:motiv_worklaods}

Prior research has focused on application-specific approaches to identify appropriate operations to be offloaded to CCM 
(see \Cref{tab:back_workloadOffloading} in \Cref{sec:background}).
Offloading the specified operations results in reduction of data movement from CXL memory to local hosts, compared to when using only the memory expansion functionality.
For example, if we run \texttt{PageRank} (i.e., graph analytics) over the expanded remote memory, the host needs to load every neighbor data per vertex on each iteration to update page rank value~\cite{famGraph}.
By offloading neighbor traversal and vertex value update to CCM, it needs to move only the updated vertex data per iteration, leaving only the page rank calculation up to the host.
In this example, the maximum data movement amount per iteration can be reduced from \{\#edge $\times$ \#vertex\} to \{\#vertex\}.

However, there is no guarantee that fixing the offloaded functions will be optimal in terms of end-to-end performance.
Depending on the input data type and the offload granularity, the offloading of the same operation, 
may shift the bottleneck to 
the host processing time or the data movement time.
We demonstrate this by running different KNN and graph analytics workloads  on multiple testbeds.

\begin{figure}[t]
    \centering
    \subfigure[Dim: 2048]{
        \label{fig:motiv_e2e_cmmAx_knn_dim2048}
        \includegraphics[width=0.452\linewidth,trim={0 0 0 0cm},clip]{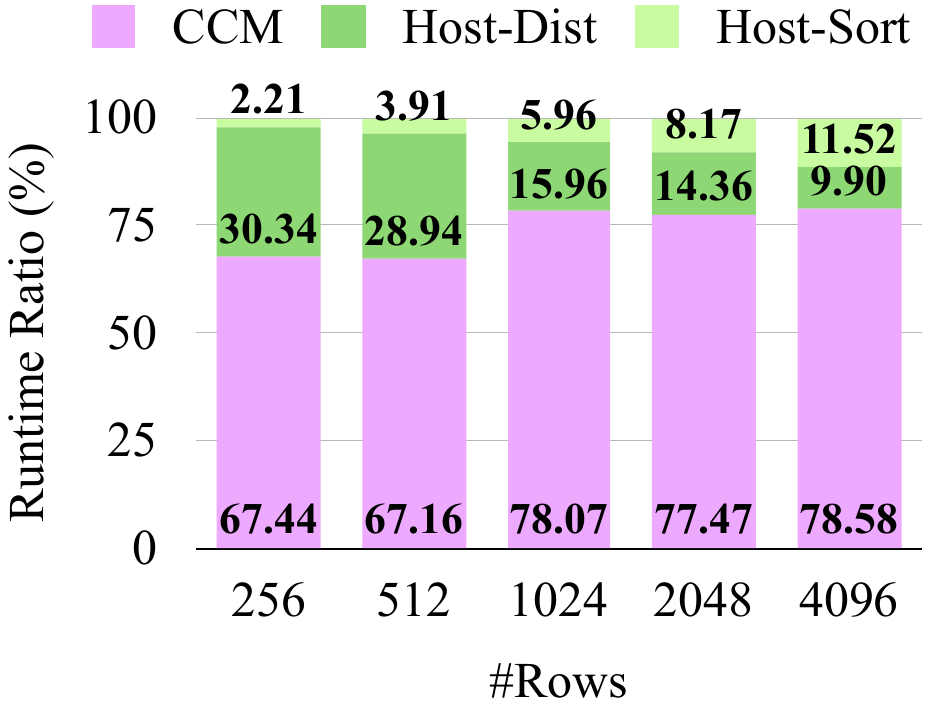}
    }
    \hfill
    \subfigure[Dim: 32]{
        \label{fig:motiv_e2e_cmmAx_knn_dim32}
        \includegraphics[width=0.452\linewidth, trim={0 0 0 0cm},clip]{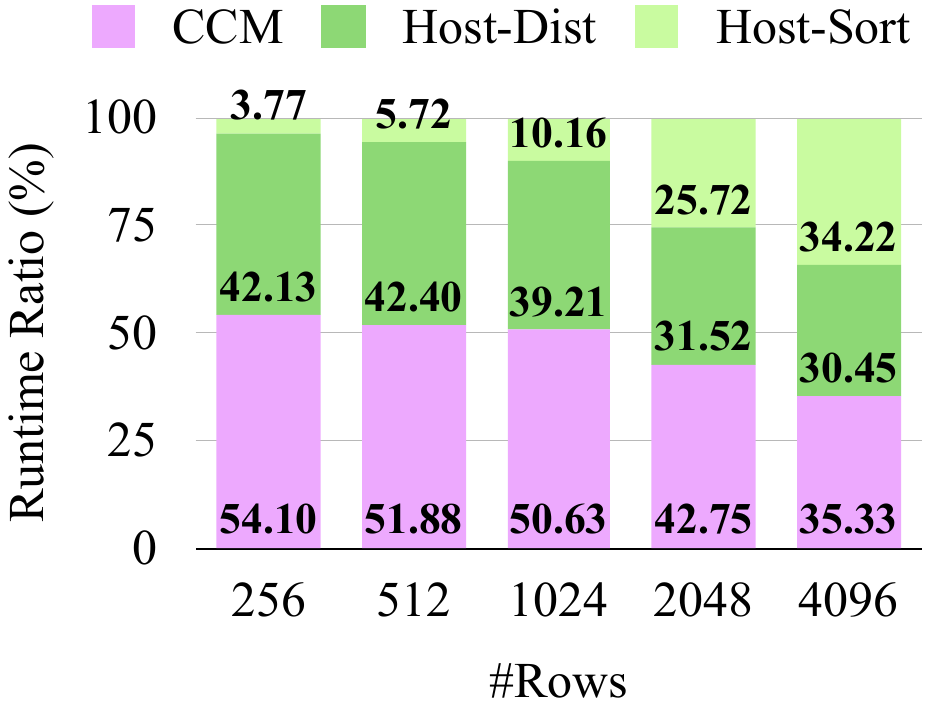}
    }
    \caption{KNN execution with various workload configurations on real hardware, showing stacked runtime ratios of CCM (purple) and host tasks (green).}
    \label{fig:motiv_e2e_cmmAx}
    \vspace{-0.15in}
\end{figure}

\begin{figure}[t]
    \centering
    \subfigure[KNNs]{
        \label{fig:motiv_e2e_m2ndp_knns}
        \includegraphics[width=0.452\linewidth,trim={0 0 0 0cm},clip]{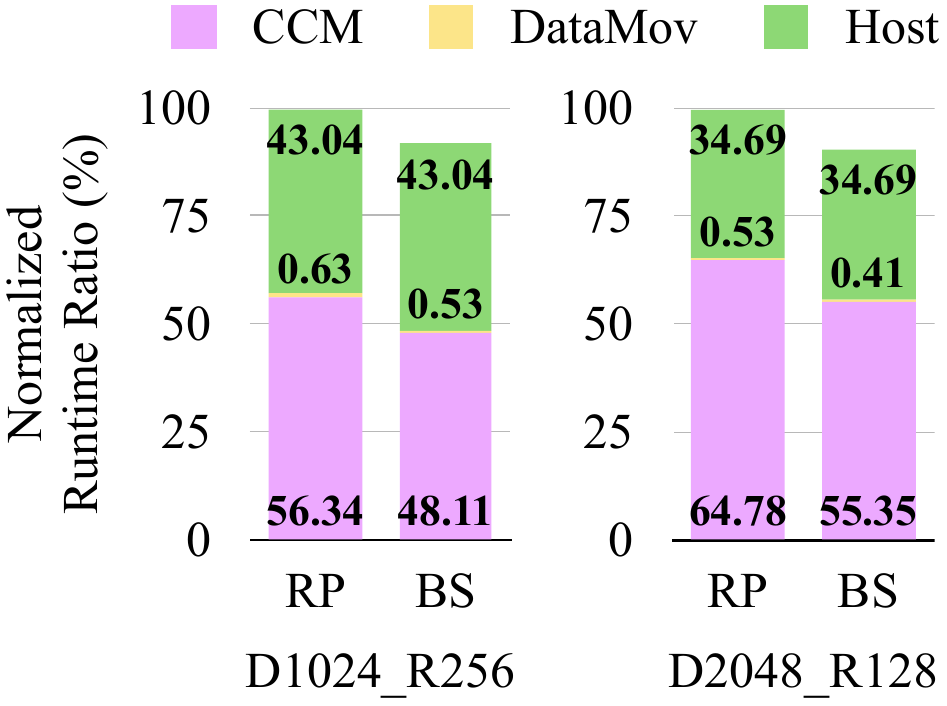}
    }
    \hfill
    \subfigure[Graph Analytics]{
        \label{fig:motiv_e2e_m2ndp_graphs}
        \includegraphics[width=0.452\linewidth, trim={0 0 0 0cm},clip]{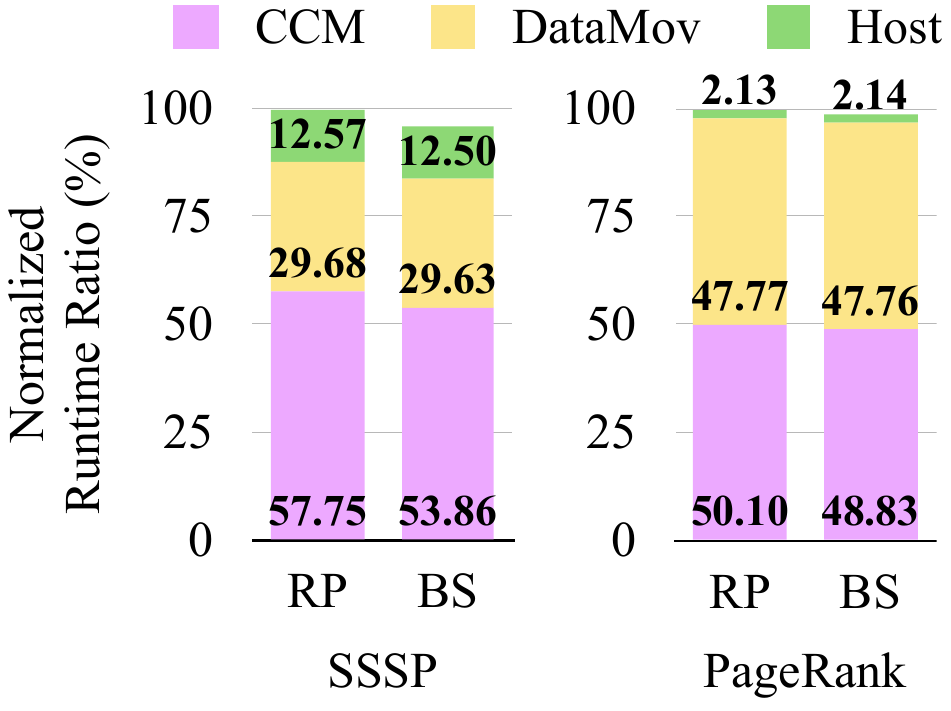}
    }
    \caption{Execution of KNNs ($D_{dim}$, $R_{numRows}$) and graph analytics on \baseline, using remote polling (RP) and bulk synchronous flow (BS) as offloading mechanisms. Normalized runtime ratios are shown as stacked bars for CCM tasks (purple), data movement (yellow), and host tasks (green).}
    \label{fig:motiv_e2e_m2ndp}
    \vspace{-0.1in}
\end{figure}

\heading{Case \#1: Host-Heavy Tasks} 
\Cref{fig:motiv_e2e_cmmAx} shows the case when running KNN for different 
vector dimension and number of input vectors in database (i.e., rows), 
 on top of the real hardware.
The graph breaks down the runtime ratio of CCM processing and host processing within the end-to-end \iscaRev{runtime}.
%
As the dimensionality decreases and the number of rows increases, 
KNN becomes a host processing-intensive application. 
Offloading vector distance calculations to CCM leads to moving a 4-byte floating point distance value per input vector to host.
The host 
receives \{\#rows\} distance values and selects the top K results.
Therefore, as the workload uses smaller dimension size per vector and more rows as input \cred{(\Cref{fig:motiv_e2e_cmmAx_knn_dim32})}, the ratio of time consumed by the host processing increases (up to 64.67\% when the dimension is 32 and the number of rows is 4096). 

Similarly, \Cref{fig:motiv_e2e_m2ndp_knns} illustrates the case where we vary the dimension size and the number of rows while running KNNs on top of the simulator, \baseline.
We simulate both offloading models, \cred{RP and BS.} 
For each workload, we normalize each time to the CCM processing time using the RP model.
The figure shows that using BS leads to a slightly shorter end-to-end runtime than using RP.
Although the CCM hardware specifications in the simulation environment differ from the real hardware (\Cref{sec:background}), the overall results indicate the same conclusion: significant host processing time, regardless of the offload mechanism.

\heading{Case \#2: Data Movement-Heavy Offloads}
\Cref{fig:motiv_e2e_m2ndp_graphs} shows a breakdown of CCM processing time, data movement time, and host processing time when running graph analytics on top of \baseline.
It shows that both the SSSP and PageRank  graph kernels result in considerable data movement time within the entire runtime.
For example, the data movement time ratio compared to the total runtime is up to \sheph{47.77\%} 
when running PageRank using the RP model.
With the increase in the number of vertices or the number of hubs (i.e. vertices with a large number of neighbors), the amount of intermediate results to be moved grows~\iscaRev{\cite{grudon}}, directly impacting the data movement time.
The increase in data sizes also puts pressure on the CXL credit-based flow control~\cite{intelCXL}, and can result in additional delays and round-trips over the CXL links.


\vspace{-0.1in}
\begin{shaded}
\heading{Observation \#2: Same offloading, Different benefit}
Common application-specific solutions focus on \textit{which} operation to offload, \sheph{but this fixed policy does} not guarantee optimal end-to-end performance, 
\sheph{as the runtime ratio} of CCM processing, data movement, and host processing 
\sheph{varies based on the workload and hardware configuration.}
\end{shaded}

\subsection{Sources of Inefficiency}\label{subsec:motiv_idleTimes}

\begin{figure}[t]
    \centering
    \includegraphics[width=\linewidth]{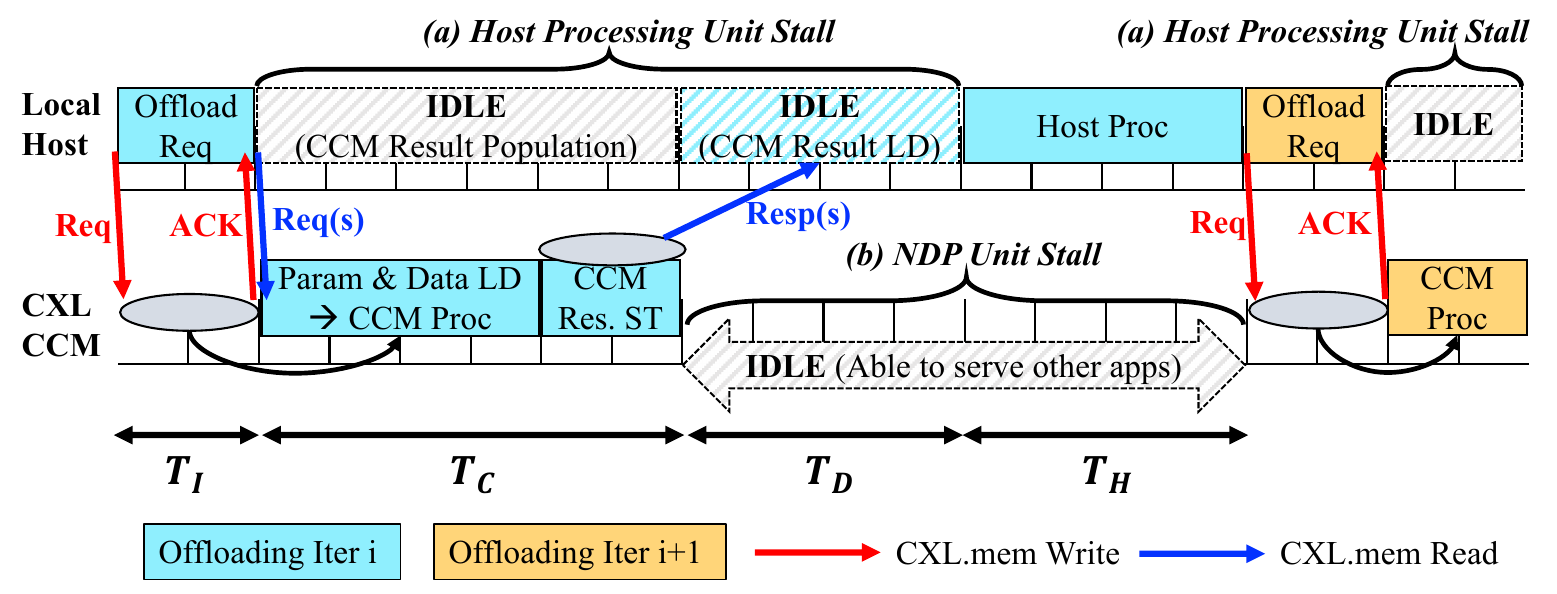}
    \caption{Naïve partial offloading based on bulk synchronous result fetch, yielding a fully serialized pipeline.}
    \label{fig:baseline_overheads}
    \vspace{-0.1in}
\end{figure}

Regardless of the offloading mechanism, the underlying host-CCM interaction relies on a CXL.mem load \cred{response} to fetch the remote processing results to local hosts.
This makes it difficult to fully utilize existing CCM systems.
\Cref{fig:baseline_overheads} illustrates how \baseline handles iterative CCM requests within a single application run.
As soon as the host receives the offload remote kernel launch ACK, it issues a CXL.mem load command
to fetch the kernel results.
With the hardware-supported barrier, the load operation is suspended until the remote kernel execution populates the final result data into remote memory.
The load command is resumed only after the CCM processing is complete, stalling the host processing unit.
This leads to significant \textit{host idle times} (\Cref{fig:baseline_overheads}(a)), equal to the CCM processing time ($T_C$) and remote result data load time ($T_D$).
\iscaRev{Additionally, partial offloading workloads exhibit dependencies across offloading requests (i.e., iterative kernels)~\cite{grudon, CXL-ANNS, neuPIMs}, indicating that} the next offload iteration may occur only after host processing is complete, \iscaRev{and host-side concurrency cannot eliminate idleness within a single application’s critical path.}
For example, in common graph analytics workloads, dependencies exist across iterations because the host must determine the new frontier based on the results of the preceding iteration.
Thus, we also observe \textit{CCM idle times} (\Cref{fig:baseline_overheads}(b)); 
the CCM module needs to wait for the next offloading iteration to launch after the result data is dispatched ($T_D$) and the host processing ($T_H$) completes.

\begin{figure}[t]
    \centering
    \subfigure[KNNs]{
        \label{fig:motiv_idleTimes_knns}
        \includegraphics[width=0.452\linewidth,trim={0 0 0 0cm},clip]{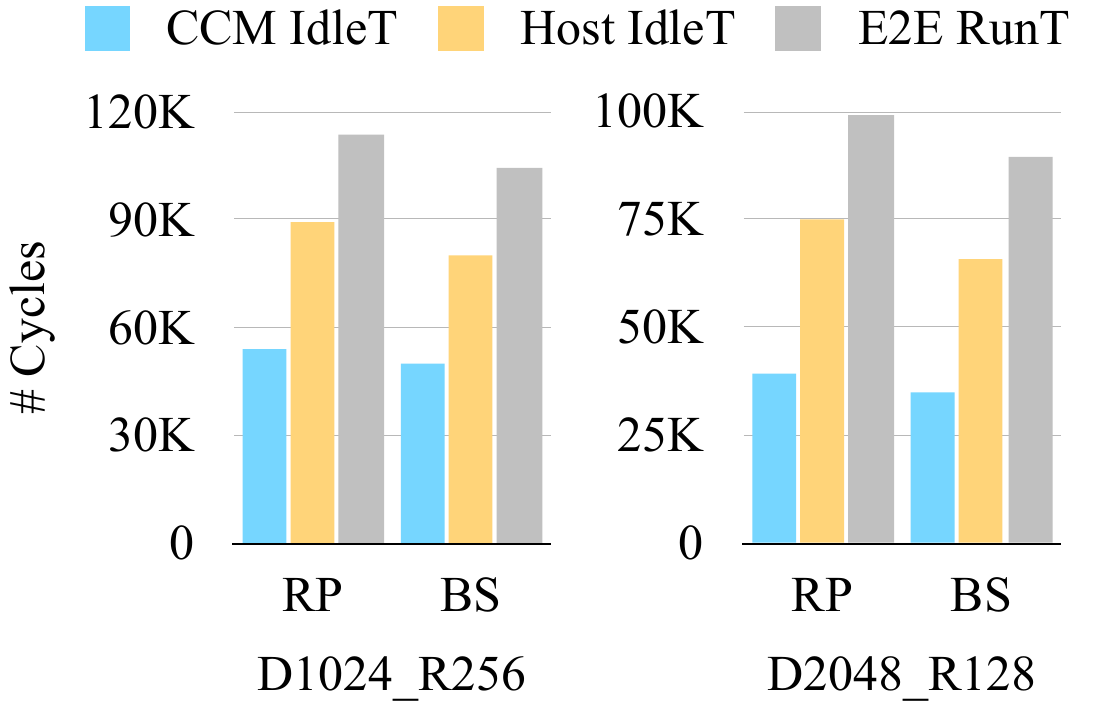}
    }
    \hfill
    \subfigure[Graph Analytics]{
        \label{fig:motiv_idleTimes_graphs}
        \includegraphics[width=0.452\linewidth, trim={0 0 0 0cm},clip]{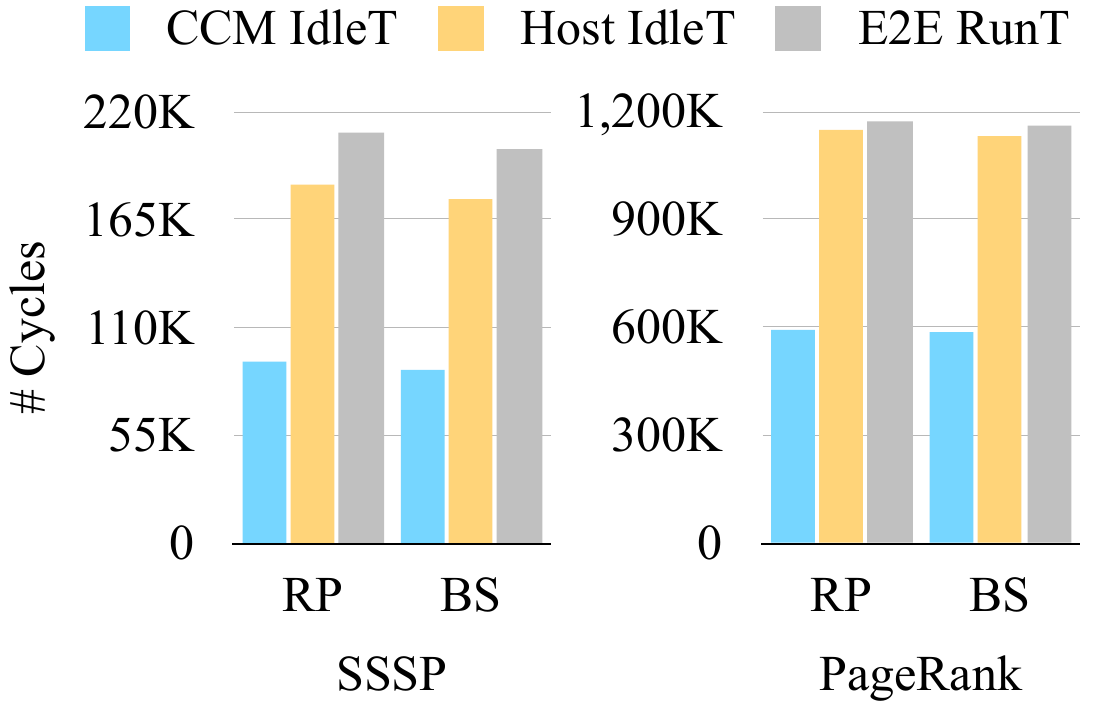}
    }
    \caption{Comparison of end-to-end runtime and two types of idle time for the same setups as in \Cref{fig:motiv_e2e_m2ndp}. Idle time is measured as the sum of task launch latency, average stall time of processing units during execution, and waiting time for task completion on the opposite side.}
    \label{fig:motiv_idleTimes}
    \vspace{-0.1in}
\end{figure}

In \Cref{fig:motiv_idleTimes}, we show the CCM idle times, host idle times, and the complete \iscaRev{runtimes} for the same workloads as in \Cref{fig:motiv_e2e_m2ndp}.
Matching time portions in \Cref{fig:motiv_e2e_m2ndp} and \Cref{fig:motiv_idleTimes} confirms high CCM idle times and host idle times in existing mechanisms.
For example, in \Cref{fig:motiv_e2e_m2ndp_graphs} PageRank on top of RP model, the runtime ratio of $T_C$, $T_D$, and $T_H$ are about 49.9\%, 48\%, and 2.1\%, 
\sheph{yielding CCM idle time ratio $\approx$ 50\% ($T_D$ + $T_H$)} 
\sheph{and host idle time ratio $\approx$ 98\% ($T_C$ + $T_D$), consistent with corresponding results in \Cref{fig:motiv_idleTimes_graphs}.}

\vspace{-0.1in}
\begin{shaded}
\heading{Observation \#3: Two idle times}
Serialized \cred{host-CCM} interaction introduces host idle time and CCM idle time, creating unnecessary bubbles in the end-to-end execution pipeline.
These idle times lower the resource utilization of the host and CCM components, limiting the usability of the general-purpose CCM systems in different scenarios. 
\end{shaded}


%% file: 04-System.tex
\section{Asynchronous Back-Streaming}\label{sec:system}

We propose a novel \textit{asynchronous back-streaming} protocol \iscaRev{for offloading in CCM systems} which can continuously overlap different components and minimize idle times in host-CCM interaction pipeline.
The main idea is to let the CXL device trigger the reverse data streaming from remote to local memory, then \cred{asynchronously} pipeline the subsequent 
data movement
to enable its overlap with CCM or host processing.
\cred{The new design is inspired by the back-invalidation snooping mechanism in the CXL.mem protocol~\cite{cxlSpec, intelCXL}, which enables a CXL device to initiate coherent memory sharing.
However, back-invalidation messages are intended to invalidate the host cache and cannot carry payloads from the CXL device to host memory.
To support \iscaRev{back-streaming} while maintaining compatibility with existing CCM models (\Cref{sec:background}), we target environments where a DMA engine is attached as a bus master on top of a CXL Type 3 device.
Further discussion on system-level implementation details is provided in \Cref{subsec:system_discuss}.}

\iscaRev{Although asynchrony and pipelining/streaming have been exploited to achieve overlap and improve performance across various domains~\cite{farm, herd, eRPC, enso, linux_uring, dpdk_ring}, enabling such high-performance communication paradigms in CCM systems requires additional non-trivial features, particularly considering trade-offs across CXL protocols (\Cref{sec:background}).
For instance, back-streaming is not part of native CXL protocols and therefore requires an additional flow control mechanism on top of transaction layers, which must be nimble and not modify the underlying CXL protocols, whereas high-performance fabrics such as RDMA can rely on hardware-supported credit management.}
We first discuss the challenges of supporting data streaming in CCM systems, followed by the design details of \system, a system that realizes low-latency asynchronous back-streaming execution based on the unmodified CXL protocol.

\subsection{Challenges of result streaming in CCM}\label{subsec:system_challenge}
To tackle the existing problems, we introduce in our protocol a mechanism to stream CCM results.
By sending the partial result data in advance, streaming allows overlap of the CCM processing time, result data load time, and host processing time.
However, it can only be useful if there is a system resolving \cred{four} main challenges between distant components: (i) how to \textit{\textbf{notify}} hosts of partial results availability, (ii) how to \textit{\textbf{expose}} result data into local region, (iii) 
how to \textit{\textbf{interface}} with \cred{concurrent executions of} CCM and host tasks without \cred{enforcing strict scheduling order or causing} stalls,
\cred{and (iv) achieving all of these while ensuring memory \textbf{\textit{correctness}}.}
Any single slowdown from these steps will result in considerable pipeline bubbles, thereby unable to solve the existing problems.

\heading{Efficient Notification} Notification of partial results is challenging due to the short result generation period, especially if the system handles fine-grained tasks.
For example, if the system stages the CCM processing for partial streaming, the single staged task can take only a single digit of microseconds scale time or even less.
Thus, the notification from remote to local is latency-sensitive and must be done \textit{rapidly} to avoid pipeline bubbles, with \textit{minimum resource usage}.

\cred{Naïve} approaches such as interrupts are not suitable since they could take up to milliseconds scale time.
Shorter polling interval (compared to the remote polling setup) is also not an option as it requires host core pinning for continuous polling over CXL link,
severely \sheph{wasting} host processing units \sheph{across multiple partial result addresses}.
Batching multiple results is available to avoid these notification overheads, however, it might result in suboptimal end-to-end performance and even similar to that of non-streaming baselines.

\heading{Rapid Data Exposal} In addition to notification, the actual data needs to be moved from remote CCM to the local host region.
As shown in \Cref{subsec:motiv_worklaods}, the data movement amount and the overhead can be significant for certain applications.
Once the host triggers the result load, the data movement is synchronized, resulting in pipeline bubbles and host idle times.
Host-triggered DMA can help prevent processing unit stalls; however, the long result loading time still remains.

\heading{Interface to Different Schedulers without Synchronization} Streaming data and pipelining 
interfaces with both the CCM and the host tasks schedulers.
It is challenging to efficiently coordinate among them since tasks are highly parallelized and each component commonly equips different schedulers.  
Those schedulers are already optimized on each CCM and host side in terms of different computing/resource capabilities, application considerations, etc.~\cite{CXL-ANNS, neuPIMs, PIFS-REC, BEACON}\iscaRev{,~\cite{grudon}}.
Thus, we need 
\cred{the host-CCM interface to integrate with}
existing parallel task schedulers, yet keeping them isolated, without 
imposing ordering or
synchronization between 
them.

\heading{\cred{Ensuring Memory Correctness}} \cred{As the hosts and the CCM device are physically separated, the system must be carefully designed to ensure memory correctness during their interactions without compromising end-to-end performance. We identify several potential issues that can arise when CCM systems fail to guarantee memory correctness.
The first is the \textit{reordering problem}, which occurs when a data or payload write precedes a flag write.
The second is the \textit{visibility problem}, which arises because the host and device are not mutually visible, allowing overwrites or unintended writes to extend beyond the fixed-size memory region.
The third is the \textit{partial write problem}, where the reader (i.e., the host) accesses data that is still being written by the device.
The fourth is the \textit{cache staleness problem}, where back-streaming updates a host memory region that has already been read and cached.}

\subsection{Overview of \system}\label{subsec:system_overview}

We design a system named \system, which integrates the new asynchronous back-streaming protocol and control plane support to effectively overcome all challenges. 

\begin{figure}[t]
    \centering
    \includegraphics[width=0.66\linewidth]{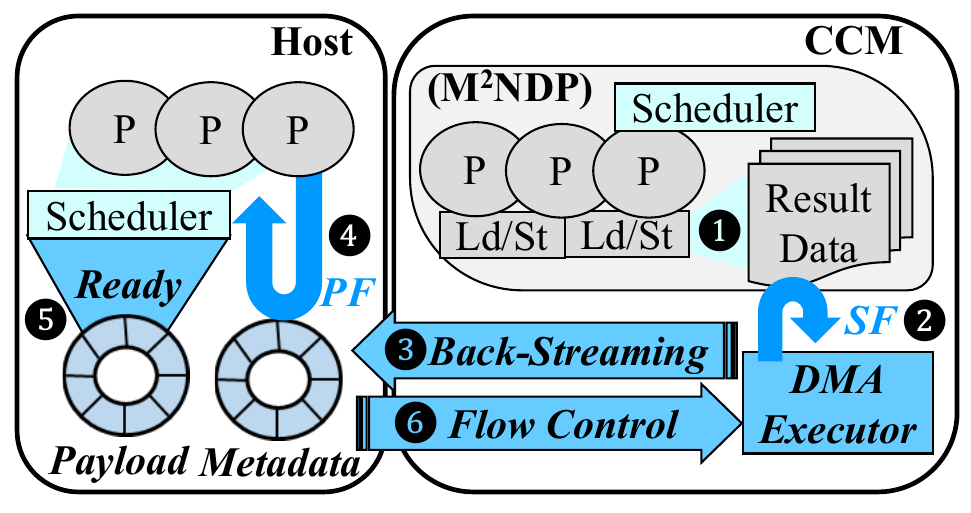}
    \caption{Overview of \system components built on top of \baseline. Dark blue shapes indicate new components in \system, while light blue shapes represent existing components interfacing with \system.}
    \label{fig:AXLE_overview}
    \vspace{-0.1in}
\end{figure}

\Cref{fig:AXLE_overview} illustrates in dark blue the overall \system components  across the host and CCM modules. 
The CCM modules adopt a fine-grained multithreaded architecture~\cite{fgmt}, as in \baseline.
It employs $\mu$threads that interleave execution by rapidly switching among one another, ensuring a steady instruction fetch, effectively hiding memory access latency and enabling high parallelism.
In \baseline, each processing unit integrates 16 $\mu$threads.
When the host offloads a task kernel, the CCM scheduler partitions the task such that each $\mu$thread processes a fixed-size input vector.
Its scheduling policy is designed to balance the load across $\mu$threads while maximizing CXL memory bandwidth utilization.
On the host side, we extend the architecture with different hardware configurations to represent general-purpose cores.
For instance, we configure two $\mu$threads per processing unit to emulate hyper-threading.

First, the host offloads the target \sheph{CCM} kernel by issuing a CXL.mem store request, 
\sheph{as} in the BS model, \sheph{but without blocking for synchronous completion.}
Multiple $\mu$threads within the CCM 
process the  store instruction and populate result data (\circledchar{1}{1}), with 
order determined by the CCM scheduler’s policy.
A DMA executor of \system monitors the result data and prepares DMA execution.
It 
forms a single \textit{payload} when continuous result data size reaches the 
DMA slot size.
\system uses ring buffers for various purposes (\Cref{subsec:system_detail}) on the local host region.
Thus, the DMA slot size \sheph{equals} 
the ring buffer slot size, which is by default 32 bytes and configurable.
The DMA executor also creates 
\textit{metadata} per 
payload.
When the pending payloads’ size gets equal or larger than the \textit{streaming factor} (\textit{SF}) (\circledchar{2}{1}), the DMA executor triggers back-streaming of payloads and metadata using CXL.io DMA (\circledchar{3}{1}).

The host has two separate ring buffers in its local DMA region for payload and metadata.
The host polls only the tail pointer of the metadata ring buffer every polling interval (\textit{PF}), which is configurable.
When the metadata tail is updated, the host knows 
new partial results have arrived in its local region (\circledchar{4}{1}).
Then, the polling routine fetches all the metadata slots that are ready, from its head \cred{index} to (tail \cred{index} - 1), and places them in the \textit{ready pool}, 
a direct interface to the host scheduler.
The host scheduler can pick the target tasks in the ready pool following its own scheduling policies.
By seeing the metadata in the pool, the host knows which payload slot to fetch to execute \sheph{downstream} 
task, where it actually loads the dependent partial CCM result data for its execution from the local region (\circledchar{5}{1}).
After processing metadata and payload ring buffer slots, the host sends flow control messages with the updated \cred{indexes} for each head to the CCM device using CXL.mem (\circledchar{6}{1}).
This ensures correct DMA region management by preventing any overwrite or overflow of the fixed size of the ring buffers. 

\sheph{After all offloading iterations complete, application completion is detected either explicitly via a tagged final CXL.io message, or implicitly once all downstream host tasks are triggered.
We adopt the latter for our single-application setting.
The former better suits for multi-tenant environments, where completion timing information must be tracked explicitly on a per-tenant basis, for example, to schedule subsequent tenants' workloads upon the completion of each offloading request.
}

\subsection{Design Details}\label{subsec:system_detail}

\begin{figure}[t]
    \centering
    \includegraphics[width=0.895\linewidth]{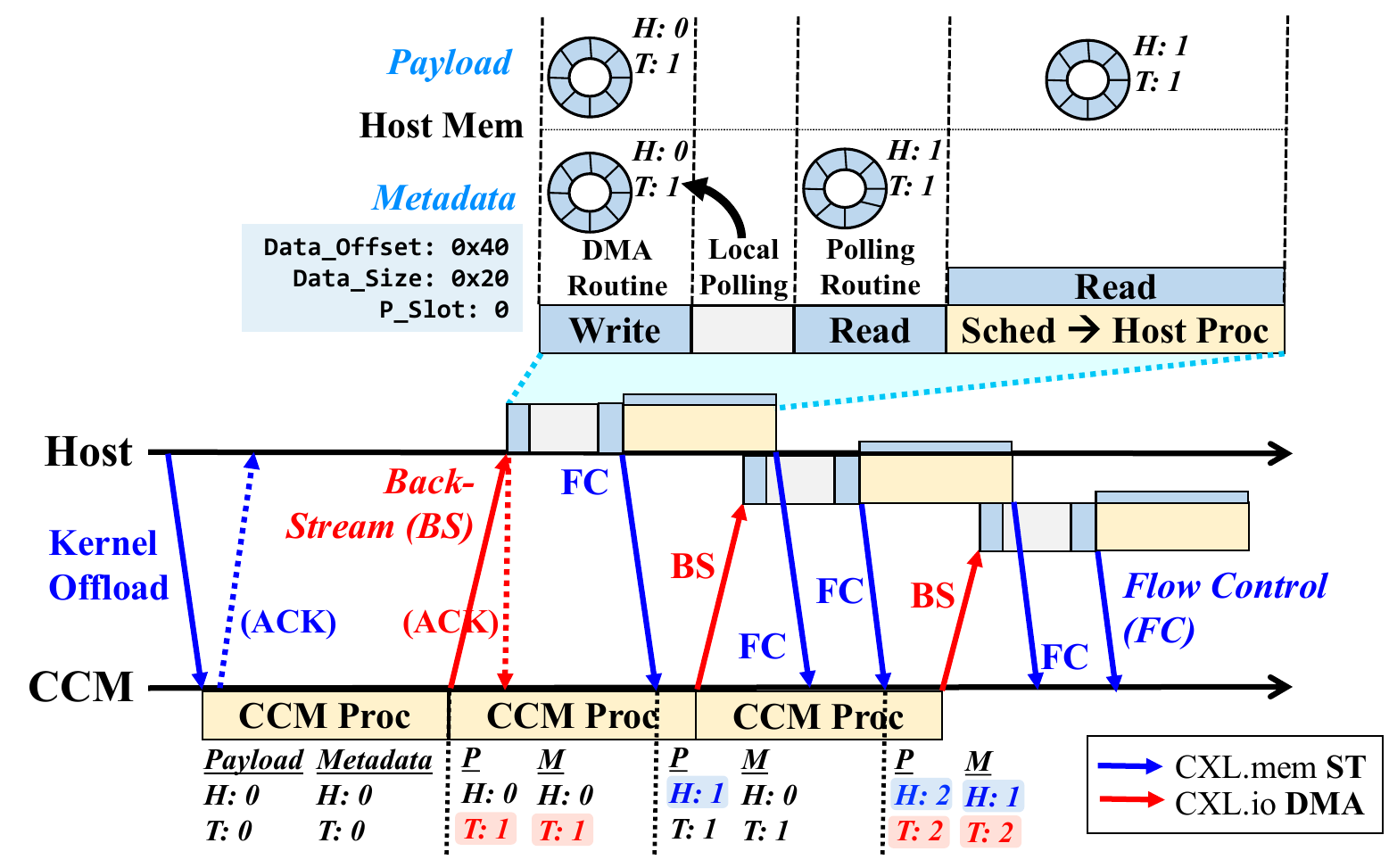}
    \caption{Detailed example flow of asynchronous back-streaming protocol and \system mechanisms. ACKs are omitted after first set of memory operations.}
    \label{fig:system_detail}
    \vspace{-0.1in}
\end{figure}

We describe the key design features and explain how each resolves the aforementioned challenges.
In \Cref{fig:system_detail}, we show the details of the asynchronous back-streaming protocol and related \system mechanism, highlighting the communication and task overlap between host and CCM modules.

\headingit{Lightweight Task Pipelining}
To enable quick and efficient notification of partial result availability, \system (a) moves the polling point into the local host region, (b) segregates the DMA region into metadata and payload ring buffers, and (c) supports complete asynchronous CCM-host communication.
This design allows the host processing unit to poll the single local address of metadata tail pointer.
\system decouples metadata consumption from payload consumption such that the polling routine only moves metadata to the ready pool (\Cref{fig:system_detail}).
Overall, the polling and its routines are  lightweight, allowing for rapid notification of partial results with minimal host resources.
In our evaluations (\Cref{sec:eval}), we demonstrate that a single-digit microseconds scale polling intervals are enough to handle even fine-grained tasks.
These numbers show that \system can deliver results 
quickly, 
while allowing the host  to switch to other tasks without being idle~\cite{shenango}.

Using ring buffers to manage DMA region allows complete asynchronous communication between distant modules, indicating that both the host and CCM continuously perform their own work without waiting for any ACKs or control messages from the remote party.
The one  key message that the asynchronous back-streaming protocol carries is a flow control message sent from host to CCM.
It is crucial to manage the local DMA regions, which are fixed-size buffers and invisible from the remote device.
Otherwise, CCM might overwrite the new data into unconsumed buffer slots or send data that overflows the total region size.
The protocol performs flow control by sending CXL.mem store operations (blue arrows in \Cref{fig:system_detail}) to alert updated payload/metadata head \cred{indexes}.
At this point, CCM \cred{no longer needs} to wait for flow control messages to update corresponding \cred{local} head \cred{indexes; it can continue processing next tasks and streaming results.} 
This is because stale \cred{CCM} head index \cred{remains} conservative enough \cred{for safe} local DMA region management.
In other words, CCM can stream data as long as its tail \cred{index} does not \cred{advance beyond} the potentially out-dated head \cred{index}. 

\headingit{Back-Streaming}
The core design of the asynchronous back-streaming protocol is to have the CCM device trigger the result \textit{send} instead of the host triggering the remote result \textit{load}.
Back-streaming transmits the partial result data in advance before the host processing units poll the notification.
Thus, when the host task is launched, host processing units can access the result data locally without any blockings.
Back-streaming is not reducing the absolute time of the result data movement, however, it allows overlapping the data movement time and the CCM/host processing times, thereby reducing end-to-end runtime 
\sheph{and freeing the host from accessing or copying remote data during its task execution.}

\headingit{\iscaRev{OoO} Streaming}
To interface with existing CCM and host parallel task schedulers, \system support \iscaRev{OoO} streaming.
This isolates the two different schedulers without the need to synchronize target tasks for pipelining.

\cred{Assume a simple scenario in which the CCM scheduler produces results in the order of data offsets 2, 0, and 1.
In this case, the result order \{2, 0, 1\} does not match the physical ring-buffer slot order \{0, 1, 2\}.
To ensure that the host processing unit retrieves the correct payload, each metadata record therefore stores the \textit{corresponding payload slot ID} separately.}
%
\cred{Now suppose the local polling routine fetches all pending metadata and places them into the ready pool.
The host scheduler may then choose to process the task associated with data offset 0 first, 
even though the earliest produced result was at offset 2.
To support such situations, the payload ring buffer operates in a \textit{gap-aware} manner, allowing non-contiguous data consumption.
The payload head index advances only after all payloads up to the maximum contiguous region have been consumed; thus, it remains at 0 even if the host has already consumed the payload in slot 1.} 


\headingit{\cred{Memory Correctness without Overhead}} \cred{By combining the data plane of asynchronous back-streaming with the control plane of \system, our design prevents memory correctness issues without introducing noticeable overhead to the end-to-end pipeline.
We describe how \system addresses each problem:
\setlength{\leftmargini}{1em}
\begin{itemize}
    \item \underline{Reordering problem:} In the current workflow, strict ordering between data writes and subsequent ring buffer tail-index updates must be maintained. Therefore, a memory fence (barrier) is required between these operations. Our simulator implementation enforces this ordering and verifies functional correctness while running applications.
    \item \underline{Visibility problem:} From the host’s perspective, the CCM is invisible, as DMA acknowledgments are returned internally to the device. Hence, a separate notification mechanism is required 
    \sheph{for} result availability, which \system provides with minimal overhead. Conversely, from the CCM’s perspective, the host’s ring-buffer capacity is unknown, which may lead to overwrites or buffer overflows. \system resolves this by maintaining local head and tail indexes within the CCM, 
    \sheph{without synchronizing host indexes,} and by employing lightweight flow control messages via CXL.mem, all without introducing stalls in the pipeline.
    \item \underline{Partial write problem:} To prevent the host from reading a partially written payload, \system enforces an additional ordering constraint between two ring-buffer items: the payload must be fully written before its corresponding metadata is updated. This ordering is guaranteed through a memory fence. In summary, \system preserves the following consistency invariant: \{payload data write $\rightarrow$ payload tail index update / metadata data write $\rightarrow$ metadata tail index update\}. The host begins reading a payload only after confirming that the metadata tail index has been updated. Even if the host observes a metadata tail index that is still being written, the enforced ordering ensures that the corresponding payload data is already complete and consistent.
    \item \underline{Cache staleness problem:} DMA regions use fixed-size ring-buffer structures, therefore, the host may access the same memory address when the buffer indexes wrap around. If the DMA region is cached, the host must flush the cache whenever it accesses that address. To eliminate this overhead, \system pins DMA regions in a cache-bypass manner (\Cref{subsec:system_discuss}). Since streamed data has no temporal locality, this design choice does not reduce performance.
\end{itemize}
These memory-correctness–related design choices allow \system to maintain ring-buffer invariants, such as index wraparound and monotonic index progression, across the two remote components, while avoiding synchronization overhead.
}

\subsection{Towards Real Systems}\label{subsec:system_discuss}

\heading{\cred{Hardware Architectures}} \cred{The CCM model is built upon the CXL Type 3 device architecture (\Cref{sec:background}).
A CXL Type 3 device allows the host to access device memory, which is sufficient to support host-initiated partial offloading.
In contrast, the asynchronous back-streaming protocol requires device-initiated \iscaRev{data} transfers.
%
%
To enable \iscaRev{back-streaming} while maintaining compatibility with existing CCM models, we target environments where a DMA engine is attached as a bus master on top of a CXL Type 3 device. In this configuration, payloads are transferred from the device to the host physical address via a CXL.io (PCIe) posted write.
We configured sufficiently long CXL.io protocol latency \sheph{for} evaluation (\Cref{subsec:eval_setup}).}

\heading{\cred{Software Stack Considerations}} 
\cred{The design of \system is currently evaluated in a simulation environment.}
Assuming access to a \cred{hardware testbed} and CXL IPs, building \system \sheph{requires no changes to} 
the underlying hardware or CXL protocol.
%
\cred{On top of this, the full-system implementation consists of three primary software components: kernel-level on the host, user-level on the host, and firmware on the device.
The kernel-level component \sheph{handles the} 
CCM device driver, 
managing host DMA memory regions, and providing
abstractions for offloading while handling the heavy lifting of low-level interactions.
For example, DMA regions must be pre-pinned so that the device can directly access host physical addresses during DMA operations.
These regions should also bypass the host cache to prevent cache staleness (\Cref{subsec:system_detail}) during frequent streaming.
Because a DMA region can be physically non-contiguous, the kernel must maintain a scatter-gather list for DMA physical regions and shadow 
the descriptors to the CCM device.}

\cred{The user-level component should ensure correct communication and protocol behavior,
such as flow control messages.
We leave the design of a programming framework and APIs that allow host applications to leverage CCM across different offloading mechanisms to future work. 
}

\cred{Finally, the CCM device firmware interfaces with the OS device driver and is responsible for the main part of asynchronous back-streaming.
It should process offloading requests, monitor CCM result population, and trigger back-streaming through the DMA executor.
The DMA executor is programmable using the shadow DMA region descriptors provided by the operating systems, allowing it to specify the source and destination addresses within the DMA routine.}

%% file: 05-Evaluation.tex
\begin{table}[t]
  \centering
  \caption{Simulation setup. The CCM configuration is based on \baseline  with 16 $\mu$threads per subcore. The host is modeled with 2 $\mu$threads per processing unit.}
  \label{tab:eval_setup}
  \begin{tabular}{>{\centering\arraybackslash}m{1cm} !{\hskip 5pt \vrule width 0.3pt} >{\raggedright\arraybackslash}p{6cm}}
    \toprule
    \textbf{Module} & \textbf{Hardware Configuration} \\
    \midrule
    Host & Processing unit \& Cache freq: 3GHz \\
                   & \# Processing units: 32, \# $\mu$Threads: 2 \\
                   & Main memory: DDR5\_4800, 16 channels \\
    \midrule
    CCM & Processing unit \& Cache freq: 2GHz \\
                     & \# Processing units: 16, \# $\mu$Threads: 16 \\
                     & CXL memory: DDR5\_4800, 16 channels \\
    \midrule
    Others & \iscaRev{Scheduling policy: Round-robin (\Cref{subsec:eval_caseStudy})} \\
                    & CXL.mem round-trip \cred{protocol} latency: 70 ns \\
                    & CXL.io round-trip \cred{protocol} latency: 350 ns \\
                    & (RP) Firmware freq: 2 GHz \\
                    & (RP) Remote polling interval: 1 $\mu$s \\
                    & \iscaRev{ (\system) Polling Interval: 50 ns, 500 ns, 5 $\mu$s} \\
                    & \iscaRev{ (\system) Streaming Factor: 32B, 64B (\Cref{subsec:eval_caseStudy})} \\
                    & (\system) Single DMA slot size: 32B, 64B \\
                    & \iscaRev{ (\system) DMA slot capacity: 50000 (\Cref{subsec:eval_caseStudy})} \\
                    & \cred{(\system) DMA preparation \iscaRev{latency}: 500 ns \iscaRev{per \sheph{req.}}} \\
                    & \cred{(\system 
                    ) Interrupt handling \sheph{latency}: 50 $\mu$s~\cite{userInterrupts} per \sheph{req.}} \\
                    
    \bottomrule
  \end{tabular}
\vspace{-0.1in}
\end{table}

\rowcolors{2}{white}{gray!10}
\begin{table}[t]
\caption{Properties of the workloads used in our evaluation.} 
\centering
\ra{.8}
\begin{tabular}{llll}
\toprule
\iscaRev{\textbf{Annot.}} & \textbf{Domain}  & \textbf{\sheph{Application}}        & \textbf{Characteristics} \\
\midrule
\iscaRev{(a)} & VectorDB    & KNN      & Dim: 2048, \#Rows: 128    \\
\iscaRev{(b)} & VectorDB    & KNN      & Dim: 1024, \#Rows: 256    \\
\iscaRev{(c)} & VectorDB    & KNN     & Dim: 512, \#Rows: 512   \\
\iscaRev{(d)} & Graph Analytics      & SSSP    & \#V: 264346, \#E: 733846\\
\iscaRev{(e)} & Graph Analytics      & PageRank    & \#V: 299067, \#E: 977676\\
\iscaRev{(f)} & OLAP   & SSB     & Query: Q1\_1~\cite{ssb}   \\
\iscaRev{(g)} & OLAP   & SSB     & Query: Q1\_2~\cite{ssb}   \\
\iscaRev{(h)} & LLM Inference   & OPT 2.7b & \#Tokens: 1K \\
\iscaRev{(i)} & \iscaRev{DLRM}   & \iscaRev{Criteo~\cite{criteo}} & \iscaRev{Dim: 256, \#Rows: 1M} \\
\bottomrule
\end{tabular}
\vspace{-0.15in}
\label{tab:eval_workloads}
\end{table}

\section{Evaluation}\label{sec:eval}

\subsection{Simulation Setup}\label{subsec:eval_setup}

\begin{figure*}[t]
    \centering
    \includegraphics[width=\textwidth]{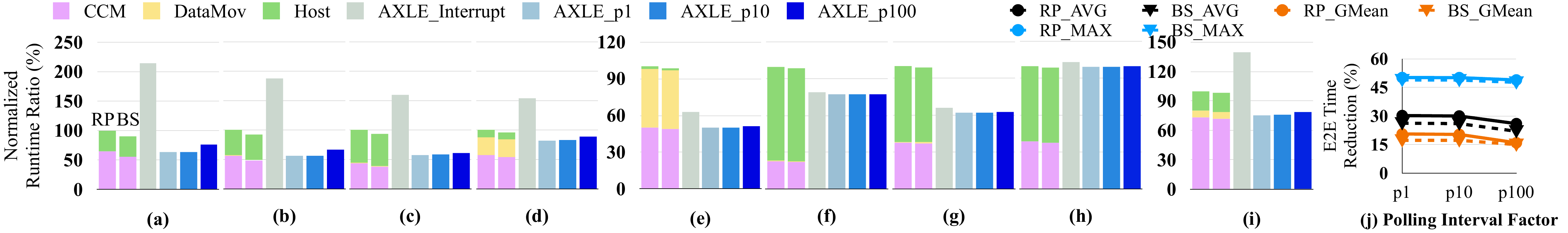}
    \vspace{-0.25in}
    \caption{\iscaRev{Normalized end-to-end runtime ratio for baselines, \system variants with interrupt-based notification, and \system (polling factors: p1 = 50 ns, p10 = 500 ns, p100 = 5 $\mu$s). (a)–(d) show lightweight tasks with fine-grained offloading, where the interrupt-handling delay becomes a severe bottleneck. (e)–(g) show longer tasks where interrupt latency is partially hidden by overlap, yet still incurs significantly higher overhead than \system using local polling.}}
    \label{fig:eval_e2eTimes}
    \vspace{-0.1in}
\end{figure*}

We implement \system on top of the open-source CCM simulator \baseline and compare it against the partial offloading mechanisms described in \Cref{sec:background}: Remote Polling (\texttt{RP}) and Bulk Synchronous flow (\texttt{BS}).
Since \baseline natively supports only bulk synchronous flows, we implement a separate RP model on top of \baseline. 
\cred{For the end-to-end runtime evaluation, we also implement an \system variant that uses interrupt-based result notification as an additional baseline (\texttt{AXLE\_Interrupt}).}
We adopt the general hardware configurations from \baseline (\cite{m2ndp}, TABLE IV), with minor modifications to reflect varying computational capabilities of the host and CCM modules (\Cref{sec:background}).
\Cref{tab:eval_setup} summarizes the configuration changes applied in our end-to-end evaluation.
\iscaRev{We follow the CXL 3.0 specification~\cite{cxlSpec} and prior documentation~\cite{intelCXL, pond} to configure the latency parameters of the CXL.mem and CXL.io protocols. In particular, \cite{intelCXL} reports that CXL.io, similar to PCIe, can exhibit a pin-to-pin round-trip latency of approximately 275 ns on Intel Xeon platforms. In our evaluation, however, we adopt a more conservative latency value.
For DMA preparation overhead, we assume a one-way control-plane latency (e.g., descriptor stores), while excluding data preparation and actual write time, which are explicitly modeled as memory operations in the simulator. Overall, our configuration remains conservative compared to recent PCIe latency measurements~\cite{pcieLatency}.
}

We evaluate \iscaRev{nine} representative workloads across \iscaRev{five} domains, following the \iscaRev{partial} offloading schemes in prior studies (\Cref{tab:back_workloadOffloading}).
Workload characteristics are summarized in \Cref{tab:eval_workloads}.
We implement several workload kernels in addition to the benchmarks used in \baseline~\cite{m2ndpSim} in a similar RISC-V instructions form.
The kernel instructions are executed by all of the $\mu$threads, each assigned a fixed-size input vector predetermined by the host and CCM schedulers.
\iscaRev{Our goal is to evaluate diverse host-CCM interaction patterns and offloading boundaries, 
defined by varying the relative combinations of CCM task length, data movement volume, and host task length.
\sheph{Accordingly, although some evaluated inputs are relatively small due to simulation constraints, our focus is on capturing relative communication–computation ratios rather than absolute dataset sizes; these trends are expected to remain consistent under scaling, and the results can be generalized to larger inputs.}
\sheph{\Cref{fig:eval_e2eTimes} shows that the  workloads in \Cref{tab:eval_workloads} represent a wide distribution of different CCM task time, data movement time, and host task time ratios.
For example, the OLAP and LLM workloads are dominated by host-side execution, while DLRM is dominated by CCM-side computation.
In VectorDB, data movement time is marginal and the remaining components are relatively balanced, whereas the Graph workloads have a large portion of data movement time.}
Consistent with \Cref{subsec:motiv_worklaods}, we \sheph{further} 
vary input sets within the same workload \sheph{(i.e., KNN)} to highlight sensitivity to parameters under different workload characteristics.}

\subsection{End-to-end \iscaRev{Runtime}}\label{subsec:eval_e2eTime}

\Cref{fig:eval_e2eTimes} compares the end-to-end runtime of RP, BS, \cred{AXLE\_Interrupt,} and \cred{the default} \system \cred{ under various local polling intervals.} 
For RP and BS, we stack the individual component times, whereas for \system we use a single bar since tasks are overlapped.
Each component runtime is normalized to the total runtime of RP, so the total ratio of RP is always 100\%, while BS shows a slightly lower value.
For instance, in \Cref{fig:eval_e2eTimes}(a), the total ratio is 100\% for RP and 90.46\% for BS.
\system further reduces the end-to-end runtime by overlapping tasks, achieving 
\iscaRev{63.41\%} in the same case.

\cred{As discussed earlier, rapid notification is crucial for the end-to-end pipeline, making a naïve interrupt-based mechanism an unsuitable design choice (\Cref{subsec:system_challenge}).
In \Cref{fig:eval_e2eTimes}, we demonstrate this by evaluating an \system variant that assumes an optimistic 50 $\mu$s~\cite{userInterrupts} interrupt-handling delay per DMA request (e.g., context switching and related costs).
\Cref{fig:eval_e2eTimes}(a)–(d)\iscaRev{, (i)} show that this delay becomes a severe bottleneck for lightweight tasks.
For example, using AXLE\_Interrupt in \Cref{fig:eval_e2eTimes}(a) results in a normalized runtime of \iscaRev{214.64\%} relative to RP.
\Cref{fig:eval_e2eTimes}(e)–(g) present longer tasks where interrupt latency is partially hidden by \system’s overlapping execution.
Nonetheless, AXLE\_Interrupt still incurs higher overhead than \system with local polling.}

Compared to RP and BS, \system consistently reduces the end-to-end runtime for most workloads, except in \Cref{fig:eval_e2eTimes}(h).
For instance, when running PageRank (\Cref{fig:eval_e2eTimes}(e)) with a 50 ns polling interval (p1), the total runtime ratio decreases by up to \iscaRev{50.14\%} and \iscaRev{48.88\%} relative to RP and BS, respectively.
In this case, increasing the polling interval has little effect.
In contrast, for relatively fine-grained tasks, the polling interval has a more pronounced impact.
For example, with KNN (\Cref{fig:eval_e2eTimes}(b)), extending the interval to 5 µs (p100) increases the runtime by \iscaRev{1.18$\times$} compared to using the 50 ns interval.

In \Cref{fig:eval_e2eTimes}\iscaRev{(j)}, we report the end-to-end time ratio reduction of \system under different polling intervals.
We present average, geomean, and maximum values across all workloads compared to each baselines.
With a short interval (p1), the average of the time ratio reductions across all workloads is \iscaRev{30.21\%} over RP and \iscaRev{26.22\%} over BS.
Extending the interval to p100 diminishes the benefit.
\iscaRev{Nevertheless,} we show that polling intervals of a few microseconds provide substantial improvements.
\iscaRev{Longer polling intervals introduce a clear trade-off between application performance and host core efficiency, which we analyze in detail in the later sections (\Cref{subsec:eval_caseStudy}).
}

Overall, when the workload is well parallelized, \system delivers predictable performance, as the longest-running component tends to overlap most of the remaining runtime.
For instance, in \Cref{fig:eval_e2eTimes}(f), the runtime ratios for BS are \iscaRev{22.24\%} for CCM processing, \iscaRev{0.58\%} for data movement, and \iscaRev{75.84\%} for host processing (totaling \iscaRev{98.66\%}).
In comparison, \system achieves an end-to-end runtime of \iscaRev{77.12\%}, indicating that host processing effectively overlaps the other components through pipelining.
\iscaRev{\system therefore provides general optimization across diverse workloads and configurations without requiring application-specific knowledge.
}

\begin{figure}[t]
    \centering
    \subfigure[\cred{LLM result with less processing units}]{
        \label{fig:eval_LLM_analysis_result}
        \includegraphics[width=0.294\linewidth, trim={0 0 0 0cm},clip]{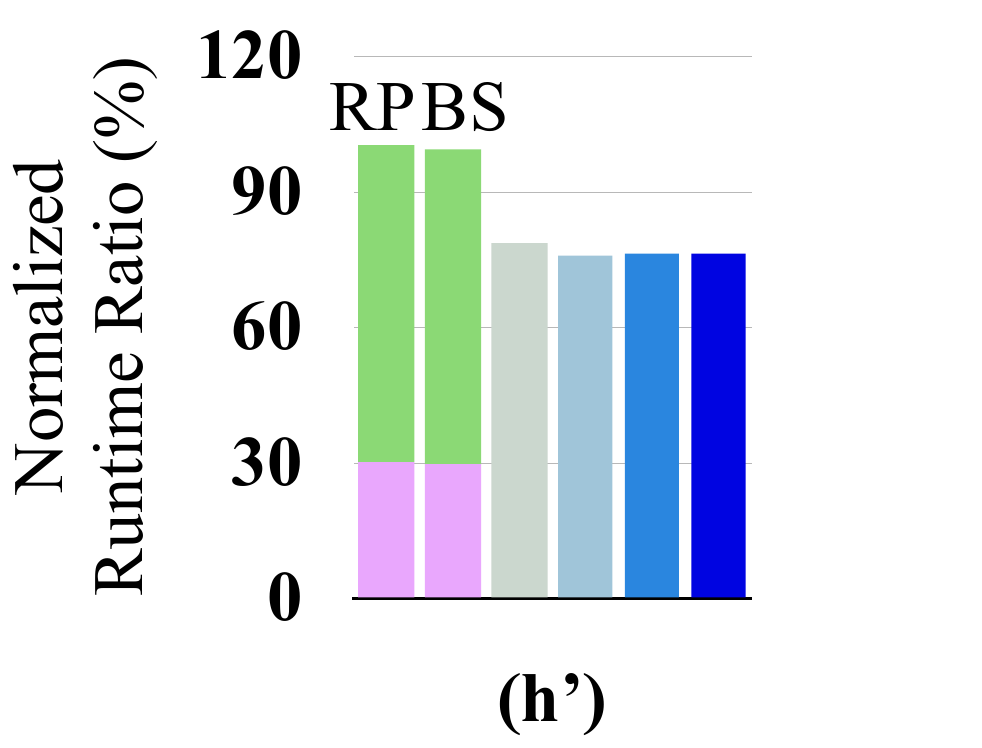}
    }
    \hfill
    \subfigure[\cred{The case of \Cref{fig:eval_e2eTimes}(h)}]{
        \label{fig:eval_LLM_analysis_16ccm_32host}
        \includegraphics[width=0.222\linewidth,trim={0 0 0 0cm},clip]{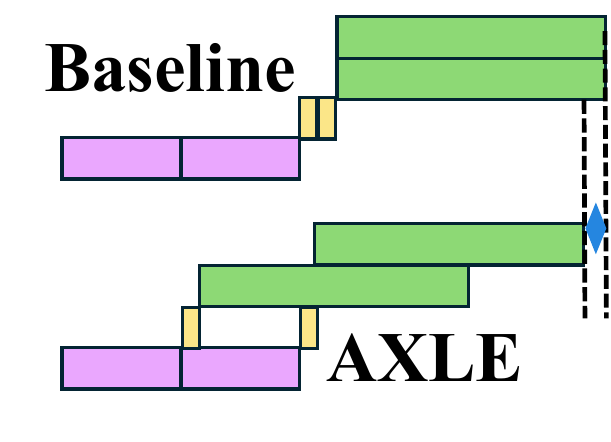}
    }
    \hfill
    \subfigure[\cred{The case of \Cref{fig:eval_LLM_anaylsis}(a)}]{
        \label{fig:eval_LLM_analysis_4ccm_8host}
        \includegraphics[width=0.334\linewidth,trim={0 0 0 0cm},clip]{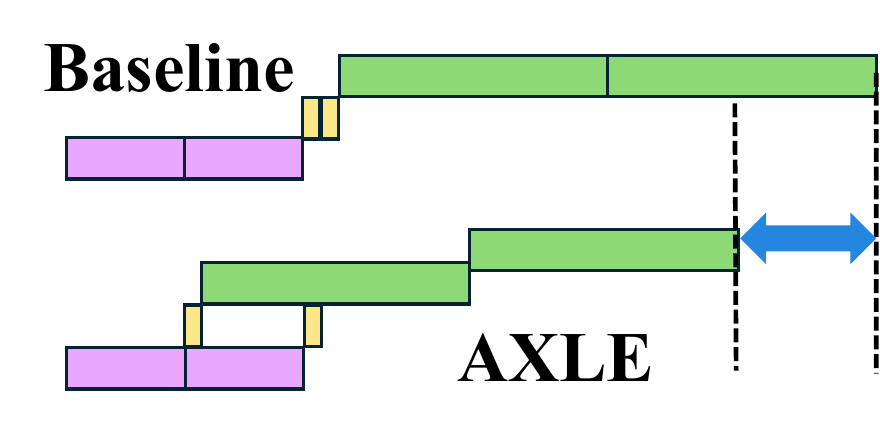}
    }
    \caption{
    \iscaRev{Different LLM-case results under modified hardware configurations: reduced processing units in both the CCM (32 $\rightarrow$ 8) and the host (16 $\rightarrow$ 4), followed by analyses of how these changes impact the end-to-end pipeline with \system. Colors and legend follow \Cref{fig:eval_e2eTimes}.}}
    \label{fig:eval_LLM_anaylsis}
    \vspace{-0.1in}
\end{figure}

\begin{figure*}[t]
    \centering
    \includegraphics[width=\textwidth]{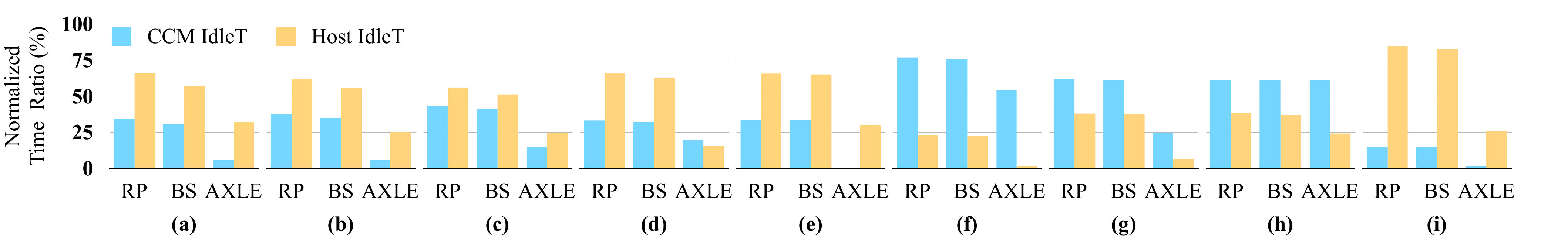}
    \vspace{-0.25in}
    \caption{\iscaRev{Normalized idle time ratio for baselines and \system when using a p10 local polling factor.}}
    \label{fig:eval_idleTimes}
    \vspace{-0.1in}
\end{figure*}

However, the performance improvement can be marginal for certain workloads \iscaRev{and configurations}, as illustrated in \Cref{fig:eval_e2eTimes}(h).
During LLM inference, the host offloads the attention block to CCM (\Cref{tab:back_workloadOffloading}), 
while the host handles the fully-connected MLP layers.
In this case, CCM processes a large dataset, but the intermediate attention output is considerably small ([1, \texttt{hidden\_size}]), \iscaRev{which leads to result sparsity}. 
\cred{As a result, 
host tasks \sheph{are far fewer} 
than 
CCM tasks due to this \iscaRev{sparse} data dependency.}
\cred{Note that even with \sheph{overlapping}, the final host task always sits at the end of the pipeline.}
When the number of host tasks is small, 
this last task\sheph{'s runtime} roughly matches the total runtime of concurrently executed host tasks in the baseline, leading to similar end-to-end performance (i.e., \cred{\Cref{fig:eval_LLM_analysis_16ccm_32host}}).
\cred{
\Cref{fig:eval_LLM_analysis_result} shows the same workload under a different hardware setup. 
With fewer host processing units, the host can no longer \sheph{batch all requests} 
\cred{(i.e., the green host tasks are no longer fully concurrent)}, making \system’s \sheph{overlap} 
more effective, as illustrated in \Cref{fig:eval_LLM_analysis_4ccm_8host}.
Consequently, \system achieves a \iscaRev{75.99\%} runtime ratio (p10) compared to RP.
}

\subsection{Two Idle Times}\label{subsec:eval_idleTime}

\Cref{fig:eval_idleTimes} shows the CCM and host idle times across workloads when running on RP, BS, and AXLE, with the local polling interval fixed to 500 ns (p10 in \Cref{fig:eval_e2eTimes}).
As discussed in \Cref{subsec:motiv_idleTimes}, idle times can be explained by aggregating the runtimes of other components.
For example, in \Cref{fig:eval_idleTimes}(f) with BS, the CCM idle time is \iscaRev{77.01\%}, which closely matches the sum of data movement and host runtime in \Cref{fig:eval_e2eTimes}(f).
Likewise, the host idle time of \iscaRev{22.99\%} aligns with the combined CCM runtime and data movement time.

\system reduces both the CCM and host idle times by overlapping component tasks, with the extent of the reduction depending on workload characteristics.
For example, in KNN with large-dimensional datasets (\Cref{fig:eval_idleTimes}(a)), the dominant CCM runtime overlaps data movement and host processing, leaving only \iscaRev{5.64\%} of CCM idle time—an \iscaRev{6.09$\times$} reduction compared to RP.
The host idle time is also halved relative to RP, but still accounts for \iscaRev{32.36\%} of total time.
This residual idle time arises because the host must wait for CCM processing to advance before streaming and pipelining intermediate results.
%
Similarly, when data movement dominates, as in graph analytics, both idle times are greatly reduced relative to RP.
However, host idle time remains non-negligible because large partial results still need to be transferred before host processing can proceed.
In \Cref{fig:eval_idleTimes}(d), \system achieves a \iscaRev{1.69$\times$} reduction in CCM idle time and a \iscaRev{4.28$\times$} reduction in host idle time compared to RP.

On the other hand, when host processing dominates, as in the OLAP case, the trend reverses.
\system minimizes host idle time, while some CCM idle time remains since it must wait for the long host execution to complete.
In \Cref{fig:eval_idleTimes}(g), \system reduces the CCM idle time by \iscaRev{2.49$\times$} and host idle time by \iscaRev{5.76$\times$} relative to RP.
As a result, the host idle time accounts for only \iscaRev{6.59\%} of the total time of the RP baseline.
\iscaRev{On average across all workloads, \system reduces CCM idle time by 13.99$\times$ and 
\sheph{13.74}$\times$ compared to RP and BS, respectively, and reduces host idle time by 3.93$\times$ and 
\sheph{3.79}$\times$.}

\iscaRev{\subsection{CCM Duality: Async. versus Sync. Execution}\label{subsec:eval_duality}}

\begin{figure}[t]
    \centering
    \includegraphics[width=\linewidth]{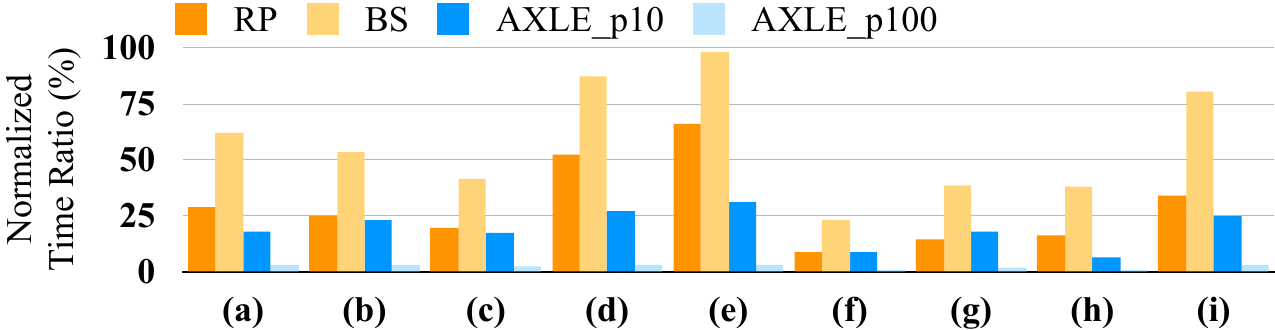}
    \caption{\iscaRev{Host core stall time normalized to end-to-end runtime across offloading cases and AXLE when using different local polling factors. Remote polling interval corresponds to p20 (1us).}}
    \label{fig:eval_host_core_stall_times}
    \vspace{-0.1in}
\end{figure}

The results in Figures~\ref{fig:eval_e2eTimes}--\ref{fig:eval_idleTimes} demonstrate that our protocol enables fine-grained offloading with \sheph{the lowest} CXL protocol overhead, achieving superior performance compared to bulk-synchronous flow.
Although the  protocol frequently uses higher-overhead CXL.io messages, it 
amortizes this cost through lightweight overlap and pipelining across system components.
In this section, we further examine the third knob (\sheph{\Cref{tab:motiv_mechnisms}}): synchronous versus asynchronous execution.

\iscaRev{\Cref{fig:eval_host_core_stall_times} shows host core stall time normalized to end-to-end runtime for each offloading case, varying AXLE's local polling interval to p10 (500 ns) and p100 (5 $\mu$s).
Host core stall time differs from previously reported idle times; earlier idle metrics are measured from the application’s perspective, whereas here we quantify how long a host core is stalled due to polling or memory operations.
The measurement includes all cycles spent on CXL (remote) and host (local) memory operations involved in host–CCM offloading interactions.
}

\iscaRev{The results show that \system significantly reduces host core stall time compared to the baselines.
In \Cref{fig:eval_host_core_stall_times}(e), host core stall time accounts for 65.99\% of total runtime for RP and 97.83\% for BS.
In contrast, it is only 30.71\% for \system with p10, a 3.19$\times$ reduction over BS.
RP polls less frequently than p10, but its remote register access incurs higher latency.
RP also relies on CXL.mem to load CCM results, counted as host stall time.
BS uses synchronous CXL.mem to offload tasks, thus PNM execution and result loading are fully counted as stall time.
\system, in contrast, offloads tasks asynchronously via CXL.mem, performs completion checks locally, and moves results with CXL.io DMA without host intervention, minimizing host core stall time across all workloads.
}

\iscaRev{With p100, \system polls local region less frequently, resulting in a single-digit ratio of host core stall time.
This configuration not only maximizes the reduction of host core stall time but also indicates that a microsecond-scale polling interval is long enough to allow processing units to perform useful work instead of spinning~\cite{shenango}.
Combined with the end-to-end runtime results (\Cref{fig:eval_e2eTimes}), p10 (or smaller intervals) can be chosen to optimize a single workload performance, whereas p100 provides a better balance between workload performance and host core efficiency in multi-tenant environments.
}

\subsection{Impact of \system Parameters}\label{subsec:eval_caseStudy}

In this section, we vary the \system \cred{ systems} configurations and explore their impact on end-to-end runtime.

\heading{Impact of Different Streaming Factors}
\begin{figure}[t]
    \centering
    \includegraphics[width=\linewidth]{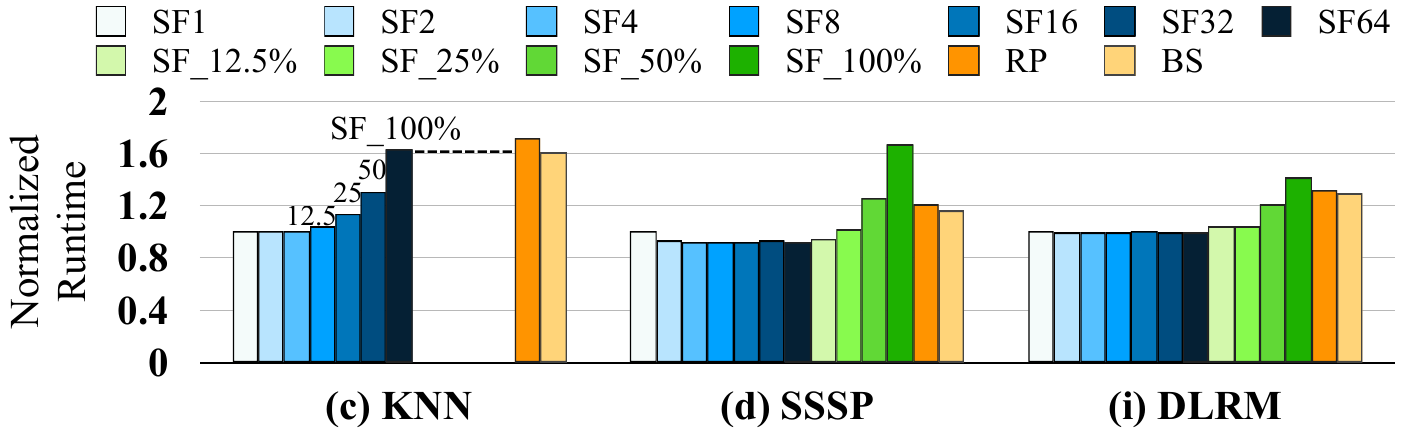}
    \caption{\iscaRev{Normalized end-to-end runtime of \system and baselines relative to SF1 across different \system streaming factors. SF$X$ (blue) denotes a streaming factor of 32 $\times X$ bytes, while SF\_$Y$\% (green) denotes $Y$\% of the total intermediate result size. Workloads with similar trends are omitted.}}
    \label{fig:eval_sf_impact}
    \vspace{-0.1in}
\end{figure}
\Cref{fig:eval_sf_impact}
shows the normalized end-to-end runtime of \system with varying streaming factors, alongside RP and BS.
Baseline (SF1) sets the smallest streaming factor to 32 bytes, meaning back-streaming is triggered whenever 32 bytes of result data are ready.
SF\textit{N} denotes \textit{N}$\times$ larger factors than SF1.

\iscaRev{In \Cref{fig:eval_sf_impact}(a), the total result data is} 2048 bytes (i.e., 512 rows * 4 bytes), \iscaRev{thereby} we test \iscaRev{from SF1 to SF64}.
Larger streaming factors batch the results, reducing overlap and pipeline efficiency.
At \iscaRev{SF64}, \system back-streams the entire result via CXL.io DMA, which is \iscaRev{slightly} slower than BS, where the entire result is fetched via CXL.mem.
\iscaRev{In \Cref{fig:eval_sf_impact}(d), increasing SF moderately reduces end-to-end runtime.
For example, SF2–SF32 achieve about 0.93$\times$ the runtime of SF1.
This improvement occurs because larger SF amortizes DMA overheads, including per-request preparation latency and the per-batch payload buffer tail-update DMA message.
However, excessively large SF values eventually degrade workload performance.
}

\iscaRev{Longer workloads are not affected by prior SF settings, as shown in \Cref{fig:eval_sf_impact}(d) and \Cref{fig:eval_sf_impact}(i).
We also evaluate very large batch sizes using SF\_$Y$\%, where a single DMA batch contains $Y$\% of the total intermediate result size.
Up to SF\_25\%, the performance impact remains marginal; for example, \Cref{fig:eval_sf_impact}(i) shows only a 1.04$\times$ runtime compared to SF1.
However, excessive SF values such as SF\_50\% and SF\_100\% can degrade performance, even relative to the baselines.
This is because \system sends a payload buffer tail-update DMA message per batch, while issuing metadata buffer tail-update DMA messages per payload.
With very large SF values, these separate DMA messages occur simultaneously, creating significant overhead on the CXL link and pipeline, especially when data movement volume is high, as in \Cref{fig:eval_sf_impact}(d).
Nevertheless, because \system minimizes per-request pipeline overheads, large SF values do not harm workload performance until a certain threshold.
Therefore, dynamically selecting an optimal SF could benefit multi-tenant environments.
}

\begin{figure}[t]
    \centering
    \includegraphics[width=\linewidth]{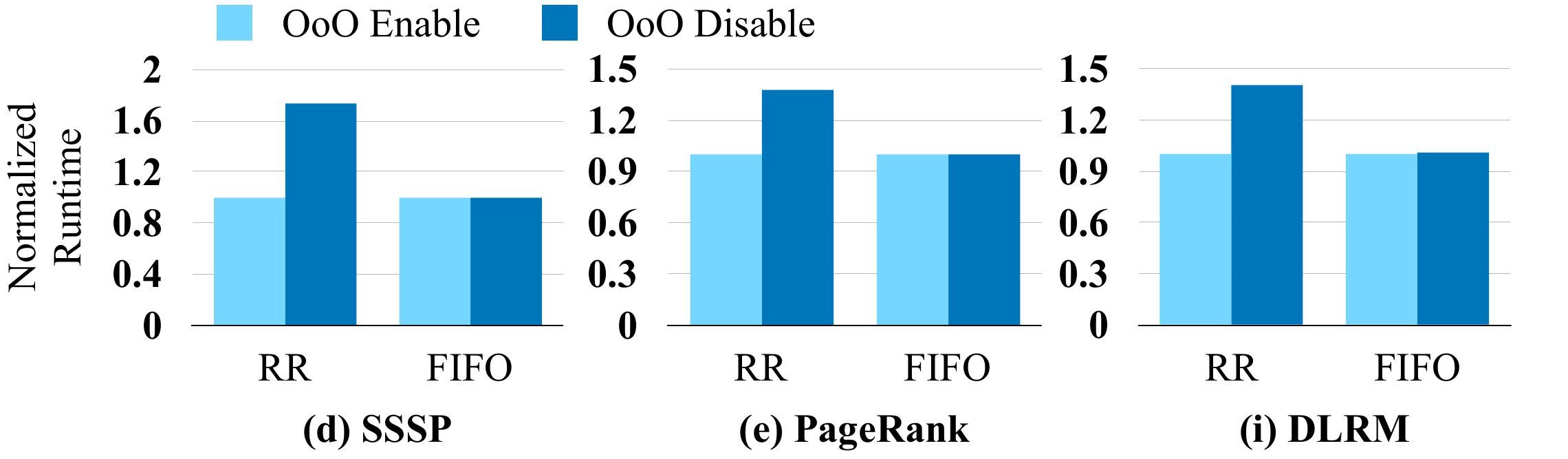}
    \caption{\iscaRev{Normalized end-to-end runtime of \system under different scheduling policies, with and without OoO streaming. Workloads for which OoO streaming does not impact the performance of a given scheduling policy are omitted.}}
    \label{fig:eval_ooo_impact}
    \vspace{-0.1in}
\end{figure}

\heading{Impact of OoO Support}
\Cref{fig:eval_ooo_impact} presents the normalized end-to-end runtime of \system under different scheduling policies, with and without \iscaRev{OoO} streaming.
Results are normalized to the case with \iscaRev{OoO} streaming enabled.
We evaluate both round-robin (RR) and FIFO scheduling, applied symmetrically to CCM and host schedulers.

By default, \system enables \iscaRev{OoO} streaming.
When disabled, the CCM enforces result ordering before transmission, triggering back-streams strictly by result offsets.
With FIFO scheduling, tasks are already processed in offset order, so enabling or disabling \iscaRev{OoO} streaming has little impact. 

In contrast, under RR scheduling, if the task at the front of the queue is not yet ready, 
it is moved to the back of the queue and the scheduler proceeds with the next available task.
With \iscaRev{OoO} streaming, \system immediately back-streams any available results, regardless of order.
Without it, the DMA executor stalls until the correctly ordered result appears, delaying transmission.
As shown in \iscaRev{\Cref{fig:eval_ooo_impact}}, disabling the feature increases runtime by \iscaRev{1.74$\times$ for (d), 1.38$\times$ for (e), and 1.41$\times$ for (i)} under RR scheduling.
These results highlight \iscaRev{OoO} streaming as a critical mechanism in \system, especially when combined with more complex scheduling policies in application-specific designs~\cite{CXL-ANNS,  neuPIMs, PIFS-REC, BEACON, grudon}. 

\heading{\iscaRev{Impact of Flow Control}}
\begin{figure}[t]
    \centering
    \subfigure[End-to-end runtime]{
        \label{fig:eval_fc_runtime}
        \includegraphics[width=0.464\linewidth,trim={0 0 0 0cm},clip]{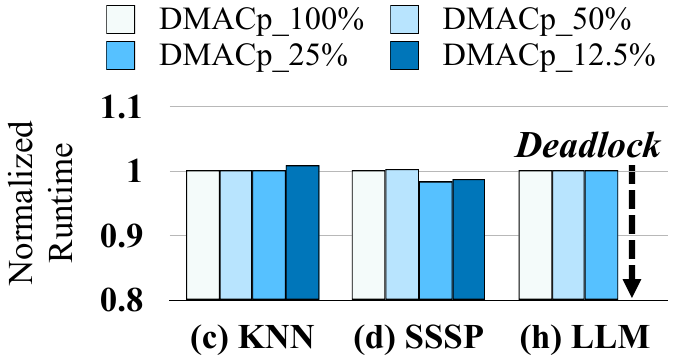}
    }
    \hfill
    \subfigure[CCM cycles waiting for credit]{
        \label{fig:eval_fc_bp}
        \includegraphics[width=0.44\linewidth, trim={0 0 0 0cm},clip]{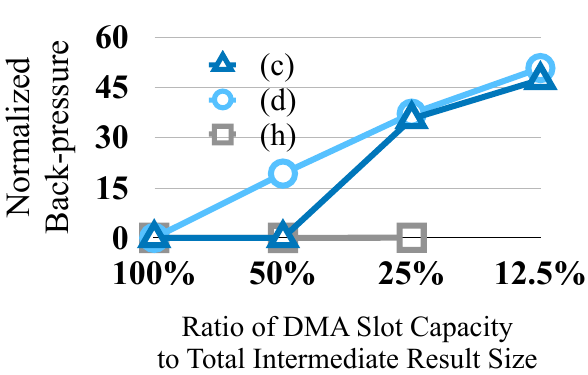}
    }
    \caption{\iscaRev{Normalized end-to-end runtime of \system under different DMA slot capacities, along with normalized back-pressure cycles (i.e., CCM waiting for credit) due to host DMA slot unavailability. Workloads whose performance matches that with abundant DMA slot capacity are omitted, where they incur zero back-pressure cycles.}}
    \label{fig:eval_fc_impact}
    \vspace{-0.1in}
\end{figure}
\iscaRev{\Cref{fig:eval_fc_runtime} presents the normalized end-to-end runtime of \system under limited DMA slot capacity, compared to the abundant configuration (DMACp\_100\%).
The results show that even with reduced DMA buffer capacity, performance degradation is marginal.
Workloads with unchanged performance across configurations are omitted; these results demonstrate that \system scales well with the number of DMA slots.
A key factor behind this scalability is the nimble flow control mechanism achieved via CXL.mem requests, making ring buffer entries quickly available after consumption.
}

\iscaRev{Another contributing factor is that \system's pipelining and overlapping effectively hide additional overhead.
\Cref{fig:eval_fc_bp} shows the normalized number of back-pressure cycles (i.e., cycles during which the CCM waits for host DMA buffer credits) relative to total runtime cycles.
The back-pressure cycles can be substantial; for example, the line corresponding to (d) (skyblue) indicates that a limited 12.5\% DMA slot capacity results in a back-pressure ratio of 50.8\% of total runtime.
Despite this, the result for (d) in \Cref{fig:eval_fc_runtime} shows that the end-to-end runtime is rather slightly reduced.
This occurs because the back-pressure impact is effectively amortized by \system's design, naturally inducing batching without additional overhead and thereby improving efficiency, consistent with the trend observed in \Cref{fig:eval_sf_impact}(d).
}

\iscaRev{Finally, (h) in \Cref{fig:eval_fc_runtime} results in deadlock when DMA slot capacity is restricted (DMACp\_12.5\%).
As described in \Cref{subsec:eval_e2eTime}, LLM exhibits sparse data dependencies between CCM and host tasks: a single host task requires sparse results from multiple CCM tasks.
Under the RR scheduler combined with \system's OoO feature, results arrive in a random order and occupy the limited DMA buffer slots, making it difficult to trigger any host task because the required set of payloads does not arrive together.
Consequently, the DMA payload buffer is never consumed, eventually leading to deadlock.
To avoid such edge cases, systems can provision sufficiently large DMA buffer capacity or employ in-order scheduling and streaming.
}

%% file: 06-Related.tex
\section{Related Works}\label{sec:related}

Prior work has developed several optimized CCM hardware solutions for different workloads, 
such as for offload of LLM inference~\cite{CXL-PNM}, approximate nearest neighbor search~\cite{COSMOS}, or entire KNN applications~\cite{CMS}. 
%
Other recent work on CCMs focuses on \textit{which} operation to offload to CCM~\cite{CLAY,PIFS-REC,m2ndp,BEACON,neuPIMs,CXL-ANNS}\iscaRev{,~\cite{grudon}}.
Their main challenge is to partition a single workload into multiple memory- and compute-intensive tasks.
For example, CLAY~\cite{CLAY} showed the benefits of offloading embedding vector/table lookup to CCM for GNN and DLRM workloads. 
\baseline~\cite{m2ndp} showed offloading boolean marking within the selection operation
can be beneficial for OLAP
workloads.
Grudon~\iscaRev{\cite{grudon}} demonstrated the benefits of 
offloading edge traversal and intermediate vertex update to CCM
to improve the performance of graph analytics. 

However, prior works overlook the question of \textit{how} to offload, and they often rely on \cred{naïve} remote polling model.
\baseline~\cite{m2ndp} is the state-of-the-art study proposing bulk synchronous flow that addresses the overhead of the remote polling model, which is why we use it as our baselines.
Furthermore, to our knowledge, \system is the first CCM system to consider the end-to-end pipeline for partial offloading, which is important for different application characteristics. 

\iscaRev{On the other hand, high-performance client-server communications have been the focus of many works across different domains~\cite{farm, herd, eRPC, enso, linux_uring, dpdk_ring}.
In these contexts, asynchrony and pipelining/streaming have been exploited to achieve overlap and improve performance.
While this work shares some high-level ideas with prior works, it provides a novel contribution in the design and evaluation of enabling such a communication paradigm in CCM systems, particularly considering \sheph{CCM duality and} trade-offs across CXL protocols (\Cref{sec:system}).
Additional features, including \sheph{efficient} notifications, use of immediate \sheph{remote} data, pipelining to downstream (host) tasks with correctness, and OoO streaming, 
are also non-trivial to identify and enable.
}



%% file: 07-Discussion.tex
\iscaRev{\section{Discussion}}\label{sec:discussion}

\heading{\iscaRev{Hardware Assumptions for CCM}}
\iscaRev{It may seem intuitive to assume that a CXL Type 2 device, which supports accelerators and CXL.cache with hardware-managed coherence, would better fit CCM by automatically synchronizing computation results with the host.
However, consistent with prior works~\cite{udon, CLAY, PIFS-REC, m2ndp, BEACON, neuPIMs, grudon, monde} and emerging industry prototypes~\cite{CMS, CXL-ANNS, COSMOS, CXL-PNM}, it is more practical to build PNM architectures on a Type 3 device, managing data and control planes via software–hardware co-design.
The primary rationale stems from CCM's core objective: achieving high memory bandwidth and capacity at low hardware cost.
In contrast to the lightweight compute logic for PNM (\Cref{sec:background}), a CXL Type 2 device requires a substantial coherence engine (DCOH), including large SRAM directories to track coherence states and complex cache-coherence logic.
These components consume more area and power than the PNM compute units, undermining memory expansion cost-efficiency.
}

\iscaRev{Furthermore, relying on CXL hardware cache coherence introduces considerable latency overheads~\cite{demystifyingCXLType2} that diminish PNM benefits.
For example, switching a memory page from Host Bias to Device Bias requires a \textit{bias flip}, forcing the host to flush caches for that page and incurring hundreds of nanoseconds to microseconds of latency.
In particular, hardware-managed cache coherence is unnecessary and only adds overhead, as CCM results are typically read-only and exhibit minimal temporal locality (\Cref{subsec:system_detail}).
Restricting the device to CXL Type 3 eliminates these hardware and software overheads, reducing complexity and cost.
}


\heading{\iscaRev{Supporting Multi-tenancy}}
\iscaRev{\system is currently focused on controlling operation offload from the perspective of individual applications.
However, we believe that 
its control plane mechanisms 
\sheph{are flexible enough to support} shared CCM use in multi-tenant environments and to support diverse resource management policies.
Future extensions of this work could address interference arising from CCM accesses, such as interconnect load caused by different SF or polling interval configurations, as well as contention for CCM resources when combining applications with long and short CCM-based computations.

}

%% file: 08-Conclusion.tex
\section{Conclusion}\label{sec:conclusion}

Existing mechanisms to offload partial operations to CCM \iscaRev{cannot leverage} 
\sheph{the underlying CXL protocols, as each treats CCM as either a device or memory alone.}
In this work, we identify those tradeoffs and demonstrate the importance of considering the host-CCM interactions from an end-to-end perspective, in order to maximize operation overlap and eliminate stalls and inefficiencies.
To support efficient general-purpose CCM systems,
this work proposes a new offloading mechanism called Asynchronous Back-Streaming, 
\sheph{which uniquely coordinates CXL.io DMA and CXL.mem to enable} efficient data streaming and \sheph{asynchronous pipelining.} 
\sheph{\system realizes this protocol with lightweight host-CCM interaction, reducing} 
end-to-end runtime by up to \iscaRev{50.14\%}, 
\sheph{application-level} CCM and host idle times \iscaRev{by an average of} \iscaRev{14.53$\times$} and \iscaRev{3.93$\times$},
\sheph{and host core stall time by up to 6$\times$.}